\newif\ifAnon\Anonfalse
\newif\ifDraft\Draftfalse

\documentclass[conference,final]{IEEEtran}
\usepackage{tugraz_defaults}
\usepackage{xcolor,colortbl}
\usepackage{subcaption}
\usepackage[frozencache=true,cachedir=minted-cache]{minted}

\graphicspath{{figures/}}

\usepackage[framemethod=tikz]{mdframed}
\mdfdefinestyle{mystyle}{%
  rightline=true,
  innerleftmargin=5,
  innerrightmargin=5,
  innertopmargin=0,
  outerlinewidth=0.5pt,
  topline=true,
  rightline=true,
  bottomline=true,
  roundcorner=1pt,
  skipabove=0.6\topskip,
  skipbelow=0.2\topskip
}

\mdfdefinestyle{tightstyle}{%
  rightline=true,
  innerleftmargin=2,
  innerrightmargin=2,
  innerbottommargin=0,
  innertopmargin=0,
  outerlinewidth=0.3pt,
  topline=true,
  rightline=true,
  bottomline=true,
  roundcorner=1pt,
  skipabove=0.6\topskip,
  skipbelow=0.2\topskip
}

\newcommand{\paragrabf}[1]{\noindent \textbf{#1}\ }

\usepgfplotslibrary{dateplot}
\usepgfplotslibrary{fillbetween}

\newcommand{\DefenseName}{Dynamic Process Isolation\xspace}

\newcommand{\CFWorkers}{\emph{Cloudflare Workers}\xspace}
\newcommand{\CF}{\emph{Cloudflare}\xspace}
\newcommand{\worker}{worker\xspace}

\newcommand{\workers}{workers\xspace}

\begin{document}

\title{Dynamic Process Isolation}

\date{}

\author{
\IEEEauthorblockN{Martin Schwarzl}
\IEEEauthorblockA{Graz University of Technology\\
martin.schwarzl@iaik.tugraz.at}
\and
\IEEEauthorblockN{Pietro Borrello}
\IEEEauthorblockA{Sapienza University of Rome\\
borrello@diag.uniroma1.it}
\and
\IEEEauthorblockN{Andreas Kogler}
\IEEEauthorblockA{Graz University of Technology\\
andreas.kogler@iaik.tugraz.at}
\and
\IEEEauthorblockN{Kenton Varda}
\IEEEauthorblockA{Cloudflare\\
kenton@cloudflare.com}
\and
\IEEEauthorblockN{Thomas Schuster}
\IEEEauthorblockA{Graz University of Technology\\
thomas.schuster@student.tugraz.at}
\and
\IEEEauthorblockN{Daniel Gruss}
\IEEEauthorblockA{Graz University of Technology\\
daniel.gruss@iaik.tugraz.at}
\and
\IEEEauthorblockN{Michael Schwarz}
\IEEEauthorblockA{CISPA Helmholtz Center for Information Security\\
michael.schwarz@cispa.saarland}}

\pagestyle{empty}

\maketitle

\begin{abstract}
In the quest for efficiency and performance, edge-computing providers eliminate isolation boundaries between tenants, such as strict process isolation, and instead let them compute in a more lightweight multi-threaded single-process design.
Edge-computing providers support a high number of tenants per machine to reduce the physical distance to customers without requiring a large number of machines. 
Isolation is provided by sandboxing mechanisms, \eg tenants can only run sandboxed V8 JavaScript code.
While this is as secure as a sandbox for software vulnerabilities, microarchitectural attacks can bypass these sandboxes.

In this paper, we show that it is possible to mount a Spectre attack on such a restricted environment, leaking secrets from co-located tenants.
\CFWorkers is one of the top three edge-computing solutions and handles millions of HTTP requests per second worldwide across tens of thousands of web sites every day.
We demonstrate a remote Spectre attack using amplification techniques in combination with a remote timing server, which is capable of leaking \SI{120}{\bit/\hour}.
This motivates our main contribution, \textit{\DefenseName}, a process-isolation mechanism that only isolates suspicious \worker scripts following a detection mechanism.
In the worst case of only false positives, \DefenseName simply degrades to process isolation.
Our proof-of-concept implementation augments a real-world cloud infrastructure framework, \CFWorkers, which is used in production at large scale.\footnote{\CFWorkers are currently used by multiple hundred thousand customers.}
With a false-positive rate of only \SI{0.61}{\percent}, we demonstrate that our solution vastly outperforms strict process isolation in terms of performance.
In our security evaluation, we show that \DefenseName statistically provides the same security guarantees as strict process isolation, fully mitigating Spectre attacks between multiple tenants.
\end{abstract}

\section{Introduction}

Cloud computing is a flexible and scalable approach to deploy applications without requiring dedicated resources.
The demand for cloud computing is still unbroken, as users benefit from the scalability, availability, security, and performance of the cloud.
The cloud provider benefits from dynamic resource allocation, maximizing usage of available resources.
To ensure high performance, cloud providers often rely on relatively modern CPUs that constantly improve the performance over the previous CPU generations. %

Many performance optimizations are done in the microarchitecture of the CPU, \ie the actual implementation of the instruction-set architecture (ISA).
Especially data-dependent optimizations, such as caches~\cite{Osvik2006,Gullasch2011,Percival2005,Yarom2014Flush} or branch predictors~\cite{Aciicmez2007e,Aciicmez2007predicting,Evtyushkin2018}, have been well-studied.
These optimizations have been shown to leak meta-data, \eg memory-access patterns of the processed data, via side channels such as timing differences~\cite{Ge2016}.
Traditionally, such microarchitectural attacks, \eg cache attacks~\cite{Yarom2014Flush,Osvik2006}, were mainly used to attack cryptographic algorithms, where the processing of secrets led to secret-dependent memory accesses~\cite{Kocher1996,Page2002,Tsunoo2003,Bernstein2005,Percival2005,Osvik2006,Zhang2012,Irazoqui2014}.
As a result, many cryptographic libraries are nowadays resilient against side-channel attacks~\cite{Bernstein2005,Coppens2009}.

With the recent discovery of transient-execution attacks~\cite{Canella2019A}, such as Spectre~\cite{Kocher2019}, Meltdown~\cite{Lipp2018meltdown}, or ZombieLoad~\cite{Schwarz2019ZL}, attackers gained a new powerful primitive.
In contrast to side-channel attacks, transient-execution attacks leak data, not meta-data.
As a result, algorithms cannot be implemented such that their secrets are not susceptible to these attacks.
As most transient-execution attacks work across logical CPUs, \ie hyperthreads, many cloud providers do not assign logical CPUs to different tenants~\cite{Weisse2018foreshadow}, reducing the attack surface to applications in the same virtual machine.

While isolating tenants through virtualization prevents exploitation of co-located tenants, it also increases the performance overhead.
Hence, cloud providers introduced edge computing~\cite{Cloudflare2019Workers,Amazon2019Lambda}, where resources are dynamically allocated by the cloud provider, on a machine that is close to the customer to ensure low latencies. 
By running the code of multiple users within the same virtual machine, the virtualization overhead is reduced, and the cloud provider's resource utilization increases.
Increasing the number of tenants per server also reduces the number of servers required at different locations. 
To still ensure isolation among tenants, cloud providers either rely on strict process isolation~\cite{Amazon2019Lambda,Microsoft2019Azure}, \ie use one process per tenant, or on language-level isolation~\cite{Cloudflare2019Workers,Fastly2021ComputeEdge,Deno2021Deply}, \ie the code has to be written in a sandboxed language such as JavaScript. 

While language-level isolation incurs the least overhead, it does not protect against Spectre.
Although microcode updates prevent mistraining of branch predictors from other processes~\cite{Intel2018white_paper}, there is no impediment for Spectre within the same process. 
Even worse, Spectre attacks have been demonstrated in JavaScript and WebAssembly~\cite{Mcilroy2019,Kocher2019,Tsuro2021SpectreJS}.
To mitigate Spectre attacks within the same process, it is necessary to insert memory barriers for each conditional branch, and avoid indirect calls and jumps.
However, these mitigations have a big impact on the performance of programs~\cite{Intel2018SpecExWhitePaper}.
Reis~\etal\cite{Reis2018siteisolation} showed that an effective mitigation in Chrome for Spectre attacks is to perform site isolation and to ensure that no untrusted code runs within the same process.
To avoid these costly countermeasures while still providing isolation among tenants, \CFWorkers rely on a modified JavaScript sandbox~\cite{Cloudflare2019Workers} that disables all known timers as well as primitives that can be abused to build timers~\cite{Schwarz2017Timers}.
This architecture allows it to support higher numbers of tenants per machine, which enables decentralized serving of applications around the world, instead of having a single centralized server per application.
\CF is one of the top three edge computing providers, \CFWorkers handles millions of HTTP requests daily.
This raises an important scientific question:

\textit{Can edge computing without strict process isolation, as is already deployed and widely used today, offer the same security levels with respect to microarchitectural attacks as edge computing with strictly isolated processes? Are the existing mitigations sufficient to prevent successful attacks?}

In this paper, we analyze the mitigations used in \CFWorkers, which remove the possibility to create a local high-resolution timer and show that they are insufficient.
\CFWorkers is a system which is kept up-to-date and relies on language-level isolation, thus we focus on microarchitectural attacks to attack it.
We demonstrate that it is possible to slowly steal secrets using an amplified Spectre attack that acquires accurate timestamps from an external time server.
However, as the resolution of the timer is not high enough, we use an amplification technique~\cite{Titzer2019Amplification} that increases the timing difference between a cache hit and a miss. 
We achieve a leakage rate of \SI{120}{\bit/\hour} inside the restricted \CFWorkers.
While this proof-of-concept attack does not pose an immediate threat, it shows that language-level isolation is not sufficient, and additional measures have to be taken. 
As already observed with other transient-execution attacks, a slow proof-of-concept implementation often just requires more engineering to make it practical~\cite{VanSchaik2019RIDL,Tsuro2021SpectreJS,Voisin2021SpectreWild}. %

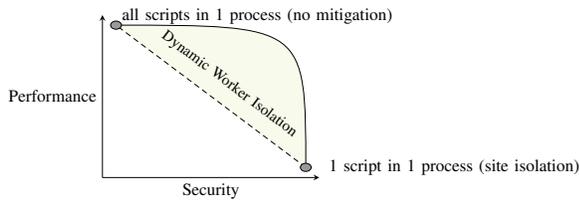
\begin{figure}
\centering
 \resizebox{0.9\hsize}{!}{
 \begin{tikzpicture}[yscale=0.75]
\small

\draw[->,>=stealth] (0,0) to node[midway,below] {Security} (4,0);
\draw[->,>=stealth] (0,0) to node[midway,left] {Performance} (0,4);

\draw[out=90,in=0,fill=green!10,looseness=1.8] (3.75,0.25) to node[midway,below,rotate=-37,xshift=-1em,yshift=-3em] {\footnotesize Dynamic Worker Isolation} (0.25,3.75);

\draw[densely dashed] (3.75,0.25) -- (0.25,3.75);
\draw[fill=black!40] (3.75,0.25) circle (0.1) node[right,xshift=1em] {1 script in 1 process (site isolation)};
\draw[fill=black!40] (0.25,3.75) circle (0.1) node[right,yshift=0.5em] {all scripts in 1 process (no mitigation)};

\end{tikzpicture}
 }
 \caption{For state-of-the-art strict process isolation, the trade-off between security and performance can be chosen by setting the number of scripts running inside the same process (dashed line).
 \DefenseName uses a detection mechanism to improve this trade-off.
 Even in the worst case, \ie a \SI{100}{\percent} false-positive rate of the detection, the performance is not worse than strict process isolation.}
 \label{fig:motivation}
\end{figure}

To prevent the exploitation of Spectre in environments such as \CFWorkers, we propose a novel technique called \textit{\DefenseName}.
This technique relies on a probabilistic detection mechanism that tries to detect ongoing Spectre attacks and isolates a suspicious workload into a separate process.
With \DefenseName we show a middle ground between the two extremes of full process isolation and language-level isolation. 
On the one hand, \DefenseName keeps the performance benefits of language-level isolation for the majority of benign workloads. 
On the other hand, \DefenseName provides the security guarantees of process isolation for malicious workloads, such as Spectre attacks. 
Even in the worst case, where every workload is classified as a Spectre attack, \DefenseName simply degrades to strict process isolation plus the small overhead of \SI{2}{\percent} for the detection, while, on average, it results in higher performance, as illustrated in \Cref{fig:motivation}.

Our detection technique relies on hardware performance counters (HPC) normalized by iTLB accesses.
We propose a novel probabilistic technique to detect Spectre attacks based on mispredicted and retired branches sampled via Intel Precise Event Based Sampling.
We show that performance counters cannot simply be used in high-performance environments as suggested by previous work~\cite{Zhang2013,Irazoqui2018mascat,Payer2016,Zhang2016CloudRadar,Chiappetta2015,Herath2015,Mushtaq2020}, as such a usage incurs non-negligible performance overheads.
However, we demonstrate that even with a limited set of performance counters, we detect running Spectre attacks when accepting a small performance overhead of \SI{2}{\percent}.

In cooperation with \CF, we evaluated our \DefenseName on a production environment in the cloud.
Even in such a diverse environment with different workloads, we see a false-positive rate of only \SI{0.61}{\percent}.
We also mounted an attack in a production environment that we detected successfully.
In all cases, \DefenseName ensures that our exploit was blocked without interrupting any of our own or other workloads.

\subheading{Contributions.} The main contributions of this work are:
\begin{compactenum}
\item We demonstrate a remote Spectre attack on the restricted \CFWorkers environment, showing that the currently deployed mitigations are not sufficient.
\item We propose a novel probabilistic detection technique for Spectre attacks with low overhead.
\item We introduce \DefenseName, a technique with, on average lower overhead than state-of-the-art strict process isolation.

\end{compactenum}

\subheading{Outline.}
The remainder of the paper is organized as follows.

In \cref{sec:background}, we provide the background.
In \cref{sec:attack}, we explain the building blocks of our attack and demonstrate a remote Spectre attack on \CFWorkers.
In \cref{sec:isolation}, we present a dynamic process isolation approach to detect Spectre attacks and dynamically isolate malicious scripts into separate processes.
In \cref{sec:evaluation}, we evaluate our approach on our production system.
In \cref{sec:discussion}, we discuss its implications before we conclude in \cref{sec:conclusion}.

\section{Background and Related Work}\label{sec:background}

In this section, we provide background on microarchitectural attacks and the current mitigations of \CFWorkers.

\subsection{Cache Attacks}\label{sec:cacheattacks}

Cache attacks exploit the access-time differences for cached and uncached data.
Since their introduction in 1996~\cite{Kocher1996}, they have been used for powerful attacks on cryptographic primitives~\cite{Kocher1996,Page2002,Tsunoo2003,Bernstein2005,Percival2005,Osvik2006}, user interactions~\cite{Schwarz2018KeyDrown,Lipp2016,Wang2019Unveiling}, or as building blocks for transient-execution attacks~\cite{Lipp2018meltdown,Kocher2019,Canella2019A}.

Nowadays, mainly two techniques are used to perform cache attacks, namely \PrimeProbe~\cite{Osvik2006} and \FlushReload~\cite{Yarom2014Flush}.
\PrimeProbe does not require shared memory and was used to create cross-core and cross-VM covert channels, attacks on co-location detection, cryptographic primitives, and SGX~\cite{Ristenpart2009,Zhang2011,Liu2015Last,Oren2015,Lipp2016,Maurice2017Hello,Schwarz2017MGX}.
\FlushReload~\cite{Yarom2014Flush} requires shared (read-only) memory between the attacker and the victim.
\FlushReload is more accurate and has been used to build local cross-core attacks on cryptographic primitives, and to spy on user behavior~\cite{Yarom2014Flush,Gruss2015Template,Guelmezoglu2015,Zhang2014,Irazoqui2015Neighbor,Irazoqui2015Lucky}.

\subsection{Transient Execution}
In modern processors, instructions are divided into multiple micro-operations (\muops)~\cite{Fog2016} that are executed out of order.
The reorder buffer ensures that the executed instructions retire in order.
Typically, a program's control flow contains conditional branch instructions.
To improve the performance of branch instructions, CPUs leverage speculative execution.
The branch prediction unit (BPU) tries to predict whether a branch will be taken or not using different data structures, \eg the Pattern History Table (PHT)~\cite{Fog2016,Kocher2019}.
Based on the prediction, the predicted branch is executed speculatively.
If the prediction was correct, the results of the execution are retired, otherwise, the speculatively executed instructions are discarded, and the correct code path is executed.
The instructions that were mistakenly executed based on branch prediction or out-of-order execution are called \textit{transient instructions}~\cite{Lipp2018meltdown,Canella2019A}.
Transient instructions still have an effect on the microarchitecture.
They can, \eg lead to measurable timing differences in the cache state.
These timing differences can then be exploited using traditional side-channel attacks to create transient-execution attacks.

\subsection{Transient-Execution Attacks \& Defenses}\label{sec:attacks-defenses}

We distinguish between Meltdown-type~\cite{Lipp2018meltdown,Canella2019A} and Spectre-type attacks~\cite{Kocher2019,Canella2019A}.
A Meltdown-type attack exploits transient execution caused by out-of-order execution after a CPU exception occured, \eg a page fault~\cite{Lipp2018meltdown}.
This enables an attacker to transiently access memory, which is architecturally inaccessible.
While the original Meltdown attack~\cite{Lipp2018meltdown} showed leakage from the cache, further Meltdown-type attacks such as Foreshadow~\cite{Vanbulck2018foreshadow,Weisse2018foreshadow}, RIDL~\cite{VanSchaik2019RIDL}, ZombieLoad~\cite{Schwarz2019ZL}, and Fallout~\cite{Canella2019Fallout} also leak data from various other internal CPU buffers.
Load Value Injection (LVI) is a transient-execution attack that turns Meltdown around and transiently injects data~\cite{Vanbulck2020lvi}.
Spectre-type type attacks~\cite{Kocher2019} exploit speculative execution.
Spectre-PHT~\cite{Canella2019A} (also known as Spectre V1) exploits the Pattern History Table, which predicts the outcome of a conditional branch~\cite{Kocher2019}.
A typical Spectre-PHT gadget is listed in~\cref{lst:basic_example}, containing a conditional branch with a bounds check.
The attacker controls the index variable \texttt{x}, which length is checked for in-bounds values.
By mistraining the branch prediction with in-bounds values, the branch will be speculatively executed for out-of-bounds indices.
The speculative execution causes a caching of the out-of-bounds accessed value.
This cached value can then be verified via a \FlushReload loop over the byte oracle (array2 in ~\cref{lst:basic_example}) to see which value was cached.
Multiple Spectre variants exist which exploit different prediction mechanisms in the CPU, for instance, the Branch Target Buffer, memory disambiguation, or the Return-Stack Buffer~\cite{Kocher2019,Kiriansky2018speculative,Horn2018spectre4,Koruyeh2018spectre5,Maisuradze2018spectre5}.
Spectre attacks were also shown by Chen ~\etal\cite{Chen2018SGXpectre} on Intel SGX.
Schwarz~\etal\cite{Schwarz2019netspectre} showed that it is possible to perform a Spectre attack over the network.
To mitigate vulnerable conditional branches, Intel and AMD propose to use serializing instructions, \ie \texttt{lfence} on both sides of the branch~\cite{IntelMitigations,AMDSpecAnalysis}.
Intel proposed several mitigations to tackle the different Spectre variants~\cite{IntelMitigations}.
Furthermore, several approaches were proposed using shadow structures for caches, adding protection domains, or to actively detect and patch Spectre gadgets in the compilation phase or with binary patching~\cite{Yan2018InvisiSpec,Kiriansky2018dawg,CarruthSLH}.

\subsection{Detection of Spectre Attacks}
Over the past decade, many cache side-channel defenses have been proposed~\cite{Brasser2019dr,Cho2020smokebomb}, with several approaches focusing on \emph{detection} of cache-based side-channel attacks using HPCs~\cite{Irazoqui2018mascat,Payer2016,Zhang2016CloudRadar,Chiappetta2015,Herath2015,Wang2020hybrid,Wang2020comprehensive,Wang2020scarf,Zhang2021see}.
To detect Spectre-type attacks, static code analysis and patching, taint tracking, symbolic execution, and also detection via HPCs were proposed~\cite{Corbet2018smatch,RedHat2018Detector,Wang2019oo7,Guarnieri2020spectector,Mushtaq2020,Gulmezoglu2019}.
Mushtaq~\etal\cite{Mushtaq2020}, Gulmezoglu~\etal\cite{Gulmezoglu2019} and Li~\etal\cite{Li2018online} propose sampling of HPC data in combination with machine learning techniques to detect attacks actively.
Related to that, Mambretti~\etal\cite{Mambretti2019} showed that HPCs can be used to analyze and debug Spectre attacks. 
However, these proposals focus on the detection of attacks, but do not propose and evaluate mechanisms to respond to protected attacks. %
As all these detection methods suffer from false positives, simply terminating a detected attack is not acceptable in the scenario of \CFWorkers.

\begin{listing}[t]
\begin{lstlisting}[caption={Spectre-PHT gadget.},label={lst:basic_example},language=C,style=customc,numbers=none]
if (x < array1_size) y = array2[array1[x] * 4096];
\end{lstlisting}
\vspace{-.8cm}
\end{listing}

\subsection{Microarchitectural Attacks in JavaScript}

Oren~\etal\cite{Oren2015} showed the first practical JavaScript-based microarchitectural attack, \ie a cache attack, that did not require installing any additional software on the victim's machine.
Kocher~\etal\cite{Kocher2019} showed that it is possible to perform Spectre attacks in JavaScript.
McIlroy~\etal\cite{Mcilroy2019} demonstrated all Spectre variants in the V8 engine.
Röttger and Janc~\etal\cite{Tsuro2021SpectreJS} demonstrated a high-performance Spectre attack in the browser exploiting L1 timers.
The main requirements for such attacks in the browser are the possibility to accurately measure time and a Spectre gadget generated by the Just-In-Time (JIT) compiler.
Since the Spectre mitigations reduce the accuracy of JavaScript timers~\cite{Mozilla2019LowResolutionTimer}, other timer techniques such as counting threads have to be used~\cite{Schwarz2017Timers,Kocher2019}.

\subsection{\CFWorkers}

\CFWorkers is a edge computing service to intercept web requests and modify their content using JavaScript.
This service handles millions of HTTP requests per second worldwide across tens of thousands of web sites.
In \CFWorkers, all \worker threads run in the same process, thus sharing the same virtual address space.
\CFWorkers support multiple thousand \workers from up to 2000 tenants running inside the same process.
Moreover, each \worker is single-threaded and stateless.
This design leads to a high-performant solution which is based on language-level isolation.
To impede microarchitectural attacks, \CFWorkers restricts the available JavaScript timing functions to only update after a request is performed.
Additionally, JavaScript worker threads are disabled, mitigating counting threads~\cite{Schwarz2017Timers,Gras2017aslr,Kocher2019}.
With those restrictions in place, they claim that it is not possible to build accurate timers.
In addition, the execution time and memory usage per \worker can be restricted.

\section{Remote Spectre Attacks on \CFWorkers}\label{sec:attack}

In this section, we show that the single-address-space design of \CFWorkers enables remote Spectre attacks.
First, we define the Spectre building blocks and overview how a remote adversary can mount a Spectre attack.
Since there is no local timing primitive, a common requirement for microarchitectural attacks~\cite{Ge2016,Schwarz2018JSZ}, we have to resort to a remote timing primitive. 
Our proof-of-concept implementation running on \CFWorkers leaks \SI{120}{\bit/\hour}. %
\subsection{Threat Model \& Attack Overview}
In our threat model, the attacker can run \CFWorkers executing JavaScript code but no native code.
Furthermore, the attacker controls a remote server to record high-resolution timestamps, \eg using \texttt{rdtsc}, and a low-latency network connection.
We also assume a powerful attacker co-located with the victim \worker, \eg by spawning multiple \CFWorkers and detecting co-location. 
As shown by Ristenpart~\etal\cite{Ristenpart2009}, an attacker spawning its instances close in time to the victim's one could maximize the probability of co-location.
\CFWorkers architecture aims to serve the same application from every location, a high number of tenants per machine is possible.
Co-location is not required for our attacker, however, this leads to the strongest possible attacker.
We assume no exploitable software bugs, \eg memory safety violations, in the JavaScript engine and no sandbox escapes.
Thus, \emph{architectural} exploits to leak data from other tenants or processes are not possible. 

\begin{figure}[t]
    \centering
\resizebox{\hsize}{!}{
 \input{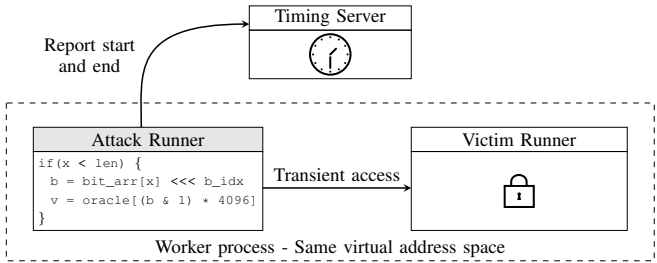}
}
    \caption{Overview of the remote Spectre attack in the \CFWorkers environment.}
    \label{fig:attack_overview}
\end{figure}

The typical requirements for state-of-the-art Spectre attacks on the timer and memory are listed in~\cref{tab:requirements}, showing the differences to our attack.
\cref{fig:attack_overview} provides an overview of our attack.
In the \CFWorkers setup, each \worker runs in the same process, and thus, shares the virtual address space.
The attacker runs a malicious JavaScript file containing a self-crafted Spectre-PHT gadget that performs a Spectre attack on its own process.
As the victim and attacker share the same process, the attacker can leak sensitive data from a victim \worker, without having to rely on an existing Spectre gadget in the victim.

Spectre attacks in JavaScript rely on speculative out-of-bounds accesses of objects. 
Assuming the attacker can either trigger a victim \worker's secret allocation, delay it, or just manages to execute before the victim, we can use heap-grooming techniques~\cite{Google2015HeapGrooming} to bring the process memory into a predictable state before both the leaking object and the victim data are allocated.
Alternatively, the attacker \worker can predict the offset between the leaking object and the victim \worker's data, or target a certain range of the virtual memory, \eg regions where V8 places similar objects~\cite{v8_release_8_3}.

For our attack, we rely on a Spectre-PHT~\cite{Kocher2019} gadget, as this is the simplest gadget to introduce in JIT-compiled code. %
Moreover, Spectre-BTB~\cite{Kocher2019} can be prevented by the JIT compiler~\cite{Google2019spectrev8}. 
In contrast to the original Spectre attack~\cite{Kocher2019}, we do not encode the data bytewise but bitwise.
The advantage of such a \textit{binary Spectre gadget} is that it is easier to distinguish two states compared to 256 states using a side channel~\cite{Bhattacharyya2019,Schwarz2019netspectre}.
While such a gadget might not be commonly \textit{found} in real applications, it is easy to \textit{introduce}.

As there are no high-resolution timers to distinguish microarchitectural states directly, we have to amplify the timing difference between a cache hit and a miss, \ie between a leaked `0' and `1' bit.
We use the amplification techniques proposed by McIlroy~\etal\cite{Mcilroy2019},and combine them with the remote measurement methods proposed by Schwarz~\etal\cite{Schwarz2019netspectre}.
With this semi-remote Spectre attack, we show that it is indeed feasible to leak data from co-located \CFWorkers in such a restricted setting.
Our Spectre attack is the only one that does not require native code execution or a local timer, or an existing gadget, and that cannot be prevented in microcode (\cf \Cref{tab:requirements}).

\subsection{Building Blocks}
As our attack uses the cache as the covert-channel part of the Spectre attack, we require building blocks for measuring the timing of cache accesses in JavaScript.
While this can be done using a high-resolution timer in some browsers~\cite{Kocher2019}, the required primitives are not available on \CFWorkers.
Hence, in addition to a different timing primitive with a lower resolution, we have to amplify the signal such that we can reliably distinguish `0' and `1' bits.

\paragrabf{Remote Timer}
On \CFWorkers, there are no local timers or known primitives to build timers~\cite{Schwarz2017Timers}.
We verified that, indeed, no technique from Schwarz~\etal\cite{Schwarz2017Timers} resulted in a timer with a resolution higher than \SI{100}{\milli\second}.
Thus, there is no possibility to accurately measure the time directly in JavaScript, and, therefore, it is not possible to perform a local Spectre attack~\cite{Kocher2019}.

\begin{table*}[t]
\setlength{\aboverulesep}{0pt}
\setlength{\belowrulesep}{0pt}
    \begin{center}
      \adjustbox{max width=\hsize}{

\begin{tabular}{l|c|c|c|r|r|r|l}
        \toprule
        \textbf{Spectre attack (variant)}                         & \textbf{Gadget}                 & \textbf{Native} & \textbf{HR Timer} & \textbf{Memory}                   & \textbf{Leakage Rate} & \textbf{Error} & \textbf{Channel}\\
        \midrule
        Kocher~\etal\cite{Kocher2019} (PHT)                      & \textcolor{red}{Yes} & \textcolor{red}{Yes} & \textcolor{red}{Yes} (ns) & \SI{2.40}{\mega\byte}    & \SI{4420.46}{\byte/\second} $\pm$ \phantom{00}\SI{6.75}{\percent} & \SI{0.07}{\percent} & Cache-L3
        \\
        Canella~\etal\cite{Canella2019A} (PHT)                    & \textcolor{red}{Yes} & \textcolor{red}{Yes} & \textcolor{red}{Yes} (ns) & \SI{3.54}{\mega\byte}    & \SI{3.13}{\byte/\second}  
        $\pm$ \SI{113.79}{\percent} & \SI{0.00}{\percent}    & Cache-L3
        \\
        Safeside~\cite{Google2019safeside} (PHT)             & \textcolor{red}{Yes} & \textcolor{red}{Yes} & \textcolor{red}{Yes} (ns) & \SI{7.00}{\mega\byte}    & \SI{4384.03}{\byte/\second} $\pm$ \phantom{00}\SI{7.75}{\percent} & \SI{0.00}{\percent}   & Cache-L3
        \\
        Canella~\etal\cite{Canella2019A} (BTB)                    & \textcolor{red}{Yes} & \textcolor{red}{Yes} & \textcolor{red}{Yes} (ns)  & \SI{6.91}{\mega\byte}    & \SI{0.71}{\byte/\second} 
        $\pm$ \phantom{00}\SI{2.43}{\percent} & \SI{0.00}{\percent}   & Cache-L3
        \\
        SafeSide~\cite{Google2019safeside} (BTB)             & \textcolor{red}{Yes} & \textcolor{red}{Yes} & \textcolor{red}{Yes} (ns) & \SI{7.01}{\mega\byte}    & \SI{269.53}{\byte/\second} 
        $\pm$ \phantom{00}\SI{0.85}{\percent} & \SI{0.00}{\percent}   & Cache-L3
        \\
        Canella~\etal\cite{Canella2019A} (STL)                    & \textcolor{red}{Yes} & \textcolor{red}{Yes} & \textcolor{red}{Yes} (ns) & \SI{3.54}{\mega\byte}    & \SI{14.37}{\byte/\second} 
        $\pm$ \SI{211.95}{\percent} & \SI{0.00}{\percent}   & Cache-L3
        \\
        Safeside~\cite{Google2019safeside} (STL)             & \textcolor{red}{Yes} & \textcolor{red}{Yes} & \textcolor{red}{Yes} (ns) & \SI{7.00}{\mega\byte}    & \SI{272.46}{\byte/\second}
        $\pm$ \phantom{00}\SI{0.22}{\percent} & \SI{0.00}{\percent}   & Cache-L3
        \\
        Canella~\etal\cite{Canella2019A} (RSB)                    & \textcolor{red}{Yes} & \textcolor{red}{Yes} & \textcolor{red}{Yes} (ns) & \SI{20.08}{\mega\byte}    & \SI{30.67}{\byte/\second}
        $\pm$ \SI{195.59}{\percent} & \SI{0.00}{\percent}   & Cache-L3
        \\
        Safeside~\cite{Google2019safeside} (RSB)              & \textcolor{red}{Yes} & \textcolor{red}{Yes} & \textcolor{red}{Yes} (ns) & \SI{7.00}{\mega\byte}     & \SI{116.70}{\byte/\second}
        $\pm$ \phantom{00}\SI{0.58}{\percent} & \SI{0.00}{\percent}   & Cache-L3
        \\
        Google~\cite{Tsuro2021SpectreJS} (PHT)               & \textcolor{green!40!black}{No} & \textcolor{green!40!black}{No} & \textcolor{red}{Yes} ($\mu$s) & \SI{15.00}{\mega\byte}      & \SI{335.02}{\byte/\second}
        $\pm$ \phantom{0}\SI{23.50}{\percent} & \SI{0.26}{\percent}   & Cache-L1
        \\
        Google~\cite{Tsuro2021SpectreJS} (PHT)               & \textcolor{green!40!black}{No} & \textcolor{green!40!black}{No} & \textcolor{red}{Yes} (ms) & \SI{15.00}{\mega\byte}      & \SI{9.46}{\byte/\second}
        $\pm$ \phantom{0}\SI{31.40}{\percent} & \SI{2.71}{\percent}   & Cache-L1
        \\
        Schwarz~\etal\cite{Schwarz2019netspectre} (PHT)           & \textcolor{red}{Yes} & \textcolor{red}{Yes} & \textcolor{green!40!black}{No} & N/A                       & \SI{7.50}{\byte/\hour} $\pm$ \phantom{000.}N/A & \SI{0.58}{\percent}   & AVX unit
        \\
        \textbf{Our work} (PHT)                     & \textbf{\textcolor{green!40!black}{No}}  & \textbf{\textcolor{green!40!black}{No}} & \textbf{\textcolor{green!40!black}{No}} & \SI{27.54}{\mega\byte}    & \SI{15.00}{\byte/\hour}
        $\pm$ \phantom{00}\SI{2.67}{\percent} & \SI{0.00}{\percent}      & Cache-L3
        \\
        \bottomrule \\
      \end{tabular}
    }
    \\
        {\footnotesize Gadget: Spectre gadget must be in victim \qquad Native: native code execution \qquad HR Timer: High-resolution timer}

  \end{center}
  
  \caption{Requirements and leakage rate of Spectre attacks. }
  \label{tab:requirements}

\end{table*}

\begin{figure}[t]
    \centering
\resizebox{\hsize}{!}{
    \input{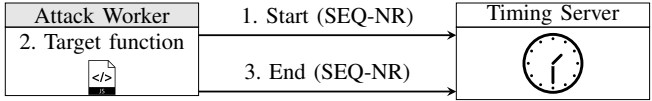}
}
    \caption{Remote timing attack on \CFWorkers. We use a high-resolution timer on an external server to measure the runtime of the target function.}
    \label{fig:attack}
\end{figure}

However, we can use requests to an attacker-controlled remote server to measure the execution time of the \worker script, as illustrated in \Cref{fig:attack}.
In this setup, the attacker sends a network request to a remote server to start a timing measurement.
The remote server stores a local high-resolution timestamp, \eg using \texttt{rdtsc}, associated with the request.
To stop the timing measurement and receive the time delta, the attacker sends another request to the remote server, which sends back the time difference from the current to the stored timestamp.
Hence, the attacker has a high-resolution time difference that is only impacted by the network latency between the attacker's \worker and the remote server.
We evaluated this timing primitive on \CFWorkers.
For the best case, \ie if the remote server runs on the same physical machine, we achieve a resolution of \SI{0.47}{\nano\second} on a \SIx{2.1}{GHz} CPU, with a jitter of \SI{1.67}{\%}.
With a resolution of \SI{0.47}{\nano\second}, it is possible to distinguish a cache hit from a miss, as required by the cache covert channel.
However, this case is not likely in reality, as the latency is typically in the microsecond range~\cite{VanGoethem2020Timeless}. 

\paragrabf{Amplification}
In our attack scenario, the attacker has no high-resolution timer but full control over the Spectre gadget.
Hence, to mount a successful attack with the remote timer, we have to rely on  amplification techniques that amplify the latency between a cache hit and miss~\cite{Mcilroy2019}.
One such technique is to transiently access multiple cache lines for a single bit instead of a single cache line and probe over these to increase the latency between a cache hit and a miss.
However, this technique is quite memory-consuming and also limited by the number of cache lines.

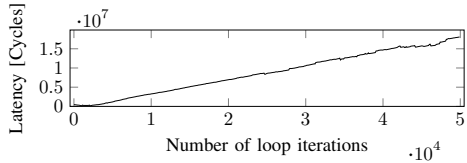
\begin{figure}[t]
    \centering
\resizebox{0.75\hsize}{!}{
    \resizebox{\hsize}{!}{
\begin{tikzpicture}
            \begin{axis}[
            enlarge x limits={0.01},
            width={\hsize},
            height=3cm,
            xlabel={Number of loop iterations},
            ylabel={Latency [Cycles]},
            xmin=0,
            xmax=50000,
            ymin=0,
            ]
            \addplot[]  table[x index = {0}, y index = {1}, col sep=comma]{data/amplification.csv};
            \end{axis}
\end{tikzpicture}
}
}
    \caption{Latency between a leaked `0' and `1' bit using an amplified Spectre-PHT attack. The latency increases with the number of loop iterations.}
    \label{fig:access_time}
\end{figure}

A way to arbitrarily amplify the latency between cache hits and misses is to either access a memory location which encodes a `0' or `1' bit transiently and then accesses the memory location for a `1' again architecturally~\cite{Titzer2019Amplification}.
\cref{lst:amplification} illustrates an \textit{arbitrary amplification}~\cite{Titzer2019Amplification} gadget.
If the Spectre gadget is optimal in terms of mistraining, we have twice as many cache misses for a `0' bit as for a `1' bit.
By using a loop over the gadget, we can create arbitrarily large timing differences between cache hits and misses.
We evaluate the amplification idea on an Intel Xeon Silver 4208, running Ubuntu 20.04 (kernel 5.4.0) in native code.
We increase the number of amplification iterations and run each iteration \SIx{1000} times to get stable results.
\cref{fig:access_time} shows the latency between a leaked `0' and `1' bit when increasing the number of amplification iterations using an amplified Spectre-PHT gadget.
As expected, there is a linear growth with the increase of the amplification iteration.
Depending on how much runtime is given to the \worker, it is possible to arbitrarily increase the delay.
Hence, we can also see that there are no strict requirements for the resolution of the remote timer. 
For lower resolutions, we can simply increase the amplification, which only results in a reduced leakage rate, but does not prevent the attack, as also shown in related work~\cite{Tsuro2021SpectreJS}.

\begin{listing}[t]
\begin{lstlisting}[caption={Amplified Spectre-PHT gadget~\cite{Mcilroy2019}.},label={lst:amplification},language=C,style=customc,numbers=none]
//transiently leak bit value
if (secret_bit) { read A; } else { read B; }
read A; //perform architectural access of 1-bit
\end{lstlisting}
\vspace{-0.6cm}
\end{listing}

\paragrabf{Eviction}
To repeat our amplification and reset the cache state, cache eviction is required.
One way to evict certain addresses from the cache is by building eviction sets~\cite{Oren2015,Gruss2016Row,Vila2019theory}.
While a targeted eviction set leads to a fast eviction, building the eviction set is costly.
Even with a local timer, the currently fastest approach takes more than \SI{100}{\milli\second}~\cite{Vila2019theory}.
In our remote scenario, this would require a lot of network requests to be performed to find the eviction set for our encoding oracle, as building the eviction set requires constant timing measurements.
Furthermore, eviction sets cannot be reused due to address-space-layout randomization on each run. %

Instead of using eviction sets, we iterate over a large eviction array (multiple MB, depending on the cache sizes of the machine) in cache-line steps (64 byte) and access the values.
If there are enough addresses accessed, the cached value is evicted~\cite{Hadad2018,Gruss2016Row,Kocher2019}.

We evaluate the eviction directly on the V8 engine used in \CFWorkers on an Intel Xeon Silver 4208, running Ubuntu 20.04 (kernel 5.4.0) as follows.
First, we access a certain index \textit{v} of a large array such that it gets cached.
Then, we perform the eviction loop and afterward verify whether \textit{v} is still cached.
To detect the optimal size of the eviction set, we increase the size of our eviction array on each iteration until we do not observe any hits \textit{v}.
For each eviction size, we repeat our measurements \SI{1000} times.
We observe that an eviction array of \SI{2}{\mega\byte} always evicts \textit{v} on our Intel Xeon Silver 4208.

\paragrabf{Bypassing Memory Randomization}\label{section:bypass_randomization}
The multi-threaded execution model of \CFWorkers uses the V8 engine to provide each \worker with an isolated environment, including a dedicated heap for JavaScript objects.
The V8 engine randomizes the location of each heap space independently.
Every pointer in each heap space is compressed, such that a pointer cannot refer to a location in a different heap space~\cite{v8_release_8_3}.

An exception is how V8 manages {\tt ArrayBuffers}: Their backing store, \ie the buffer that contains the data, is allocated in the thread's native heap.
The backing store is referred to using non-compressed pointers~\cite{v8_release_8_3}.
Therefore, the offset between two {\tt ArrayBuffers} can be predicted and influenced by classic heap-grooming techniques~\cite{Google2015HeapGrooming}.
Even if the two buffers are allocated in two different thread heap areas, their relative offset is constant across executions.
We verified this experimentally.
We thus leveraged the predictability of such allocations to leak data across \CFWorkers' {\tt ArrayBuffers}.
We evaluated the heap layout given a fixed sequence of {\tt ArrayBuffers} allocations by inspecting the V8 process memory, and confirmed that the offset is completely deterministic as long as the attacker ensures that the victim \worker maintains the same allocation sequence.

However, such a Spectre attack is not limited to {\tt ArrayBuffers}. 
As the index to the {\tt ArrayBuffer} is attacker-controlled, an attacker can provide any value to target an arbitrary virtual address in the address space. 
In contrast to memory-corruption attacks, an attacker can use Spectre to safely probe the address space, as invalid accessses do not raise an exception~\cite{Kocher2019,Goktas2020blind}. 
Göktas~\etal\cite{Goktas2020blind} introduced the concept of a speculative probing primitive that leverages Spectre to break classical and fine-grained address space layout randomization. 
Gras~\etal\cite{Gras2017aslr}, Schwarz~\etal\cite{Schwarz2019STL}, and Lipp~\etal\cite{Lipp2020takeaway} also demonstrated that microarchitectural attacks in JavaScript can break memory randomization.
Hence, memory randomization is only a small obstacle that can be deterministically circumvented using engineering.

\subsection{Attack on \CFWorkers}\label{sec:attack_eval}

Using the discussed building blocks, we mount an attack on \CFWorkers to extract secret bits from a \worker under the assumption that the location of  the targeted byte is known to estimate the best possible attack.
We mount our remote JavaScript attack as follows.
\begin{compactenum}
\item We send an initial request with a sequence number to a timing server.
The timing server stores the current local, high-resolution timestamp when receiving the request.
\item We perform a Spectre attack on a target address.
\item We send another request to the timing server.
The timing server calculates the delta between the current and the stored timestamp to distinguish between a cache hit or miss.
\end{compactenum}

The timing server infers the cache state from the delta between timestamps, \ie if two requests with the same sequence number have a small delta, a cache hit was observed, otherwise, a cache miss was observed.
Note that the cache state corresponds to the leaked secret bit.
To initially get the timing difference between an amplified `0' and `1' bit, the attacker can setup an experiment where the JavaScript architecturally produces two cache hits (`1' bit) versus a cache hit and a miss (`0' bit).
Based on the delta, the attacker chooses a threshold to distinguish the bit.
Thus, the remote timing server infers the secret bit from the timings.
As the attacker controls both the attacking \worker and the timing server, there is no need to send the leaked information back to the \worker.

There are different challenges when creating a JavaScript Spectre PoC, as the V8 JIT compiler tries to emit optimized code for a function. %
Turbofan, the V8 optimizing compiler, uses assumptions gathered from the V8 interpreter to emit optimized code tailored to the observed inputs.
If such assumptions are invalidated, the function is de-optimized.
Moreover, during the garbage collection phase, objects are moved between different heap spaces of the same \worker to reduce the memory footprint of the code.
Such dynamic code changes impact the success rate of a PoC.

\paragrabf{Evaluation.}
To develop and evaluate a proof-of-concept attack, we obtained a local developer copy of \CFWorkers to not interfere with any \worker of other customers.
We ensured that the configuration on our local system is identical to the configuration running on the cloud.
As \CFWorkers mostly use server CPUs, we also focus our attack on an Intel server CPU, specifically an Intel Xeon Silver 4208, running Ubuntu 20.04 (kernel 5.4.0).
To speed up exploit development, we do not set the \texttt{--untrusted-code-mitigations} runtime flag, which enables array index masking in V8~\cite{v8_api}.
Previous work already showed that these mitigations can be partially circumvented with the use of large and small JavaScript objects~\cite{Hadad2018}.
Furthermore, the developers of V8 found no mitigation to mitigate Spectre V4~\cite{Mcilroy2019,Titzer2019Amplification}.

We create a Spectre-PHT PoC that leaks bits from a victim {\tt ArrayBuffer} by transiently reading out-of-bounds. %
As discussed in Section~\ref{section:bypass_randomization}, using {\tt ArrayBuffer}s allows an attacker to statically compute the relative offset at which the secret is located.
We describe the optimizations to increase the leakage rate in~\cref{sec:appendix:optimization_prevention}.

We call the function performing a Spectre attack \SIx{10000} times and repeat the experiment \SIx{1000} times, observing a success rate of \SI{1}{\percent} ($n=1000$, $\sigma=1.25\%$) for an unoptimized version in terms of successfully fetching an out-of-bounds value into the cache.
We repeat the same experiment with our optimized version (\cf~\cref{sec:appendix:optimization_prevention}) and observe a success rate of \SI{54.31}{\percent} ($n=1000$, $\sigma=23.16\%$).
We assume that the attacker is capable of creating a completely stable exploit with \SI{100}{\percent} success rate.
From now on, we evaluate our metrics with a \SI{100}{\percent} success rate to estimate the best possible attack, where the attacker knows where the secret array is located.

According to NetSpectre~\cite{Schwarz2019netspectre}, to clearly distinguish between cached and uncached data on a local network, \SIx{1000000} measurements (1 measurement is the difference between two timing requests) are required.
However, if amplification is used, the number of required requests can be reduced.
We evaluate a set of different amplification factors (number of loop iterations) in native code between \SI{1} and \SI{1000}, and sample each loop length \SIx{100000}.
We implement the box test~\cite{Crosby2009} to determine the number of required requests~\cite{VanGoethem2020Timeless,Brumley2005remote,Crosby2009}.
Based on the sampled data, we create histograms and choose the threshold.
We determine the required number of requests to achieve the maximum success rate of \SI{100}{\percent}.

\begin{figure}[t]
    \centering
\resizebox{\hsize}{!}{
    \resizebox{\hsize}{!}{
\begin{tikzpicture}
            \begin{axis}[
            enlarge x limits={0.01},
            width={\hsize},
            height=3cm,
            xlabel={Number of requests},
            ylabel={Success rate},
            xmode=log,
            ymajorgrids=true,
            legend style={font=\tiny},
            legend pos=outer north east,
            grid style=dashed
            ]

            \addplot[red,mark=x]  table[x=requests,y=succ, col sep=comma]{data/success_over_requests_1k_100.csv};
            \addlegendentry{$A_{1000}$}
            \addplot[green,mark=*]  table[x=requests,y=succ, col sep=comma]{data/success_over_requests_100_1k.csv};
            \addlegendentry{$A_{100}$}
            \addplot[blue,mark=diamond]  table[x=requests,y=succ, col sep=comma]{data/success_over_requests_10_10k.csv};
            \addlegendentry{$A_{10}$}
            \addplot[orange,mark=o]  table[x=requests,y=succ, col sep=comma]{data/success_over_requests_5_20k.csv};
            \addlegendentry{$A_{5}$}
            \end{axis}
\end{tikzpicture}
}
}
    \caption{Success rate over the number of requests per number of amplification factor.}
    \label{fig:success_over_requests}
\end{figure}
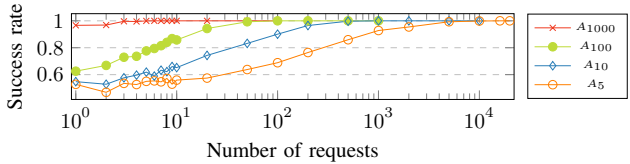

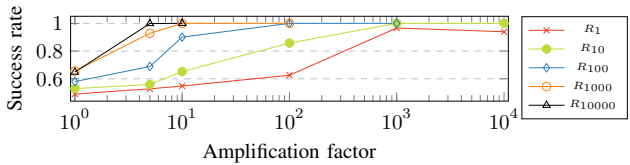
\begin{figure}[t]
    \centering
\resizebox{\hsize}{!}{
    \resizebox{\hsize}{!}{
\begin{tikzpicture}
            \begin{axis}[
            enlarge x limits={0.01},
            width={\hsize},
            height=3cm,
            xlabel={Amplification factor},
            ylabel={Success rate},
            xmode=log,
            ymajorgrids=true,
            legend style={font=\tiny},
            legend pos=outer north east,
            grid style=dashed
            ]

            \addplot[red,mark=x]  table[x=ampl,y=succ, col sep=comma]{data/success_over_ampl_1.csv};
            \addlegendentry{$R_{1}$}

            \addplot[green,mark=*]  table[x=ampl,y=succ, col sep=comma]{data/success_over_ampl_10.csv};
            \addlegendentry{$R_{10}$}

            \addplot[blue,mark=diamond]  table[x=ampl,y=succ, col sep=comma]{data/success_over_ampl_100.csv};
            \addlegendentry{$R_{100}$}

            \addplot[orange,mark=o]  table[x=ampl,y=succ, col sep=comma]{data/success_over_ampl_1000.csv};
            \addlegendentry{$R_{1000}$}

            \addplot[black,mark=triangle]  table[x=ampl,y=succ, col sep=comma]{data/success_over_ampl_10k.csv};
            \addlegendentry{$R_{10000}$}

            \end{axis}
\end{tikzpicture}
}
}
    \caption{Success rate of different amplification factors for different number of requests.}
    \label{fig:success_over_ampl}
\end{figure}

\cref{fig:success_over_requests} illustrates the number of requests required to achieve a certain success rate for different amplification factors.
As can be seen, the higher the amplification factor is, the fewer requests are required to achieve high success rates.
We are also in a similar range for the success rate of a sequential timing attack on the localhost~\cite{VanGoethem2020Timeless}.
In addition, we evaluate the success rate of different amplification factors based on a different number of requests.
As ~\cref{fig:success_over_ampl} illustrates, with small amplification factors but enough requests, we can also achieve a high success rate of more than \SI{95}{\percent}.
We refer to the work of Van Goethem~\etal\cite{VanGoethem2020Timeless} and Schwarz~\etal\cite{Schwarz2019netspectre} for the required requests in a network with multiple hops.

We evaluate our attack locally, \ie with a timing server on the same machine.
We first evaluate an optimal attack in native code.
The optimal attacker would choose the number with the highest success rate and the lowest number of requests required, resulting in the lowest execution time.

We choose a random 16-bit secret.
As amplification factor, we choose \SIx{100000} loop iterations and perform just one request.
With this setup, leaking one bit takes on average \SI{2.5}{\second} ($n=100$, $\sigma_{\bar{x}}=0.05\,\%$).
We repeat the experiment $100$ times and observe a leakage rate of \SI{23}{\bit/\second} ($n=100$, $\sigma_{\bar{x}}=2.8\,\%$).
Using an outlier filter, this error can be reduced towards $0$.
We measure an average distance between the means of the `1's and `0's of \SI{3434697} cycles.
As these values are from a native-code attack, we consider these numbers as the maximum achievable leakage rate for JavaScript.
The reason for that is that a JavaScript attacker is more restricted in terms of evicting certain addresses from the cache and thus requires additional time for the eviction.
Furthermore, the code is JIT compiled, and thus, the attacker requires a warmup to stabilize the JIT-compiled code before performing the attack.

\begin{figure}
\resizebox{\hsize}{!}{
 \begin{tikzpicture}
            \begin{axis}[
            scatter,
            style={font=\footnotesize},
            width={\hsize},
            height=3cm,
            xlabel={Bit Number},
            ylabel={Timing difference},
            ymin=69600089700,
            ymax=70702499999,
            xtick={1,...,16},
            xticklabels={1,0,0,0,1,0,0,1,0,1,1,0,0,1,1,1},
            scatter/classes={%
                a={mark=square*,blue},%
                b={mark=triangle*,red}
            },
            ymajorgrids=true,
            legend style={font=\tiny},
            legend pos=outer north east,
            ymajorgrids=true,
            grid style=dashed
            ]

            \addplot[scatter,only marks,mark=*,scatter src=explicit symbolic]  table[x=Number,y=Delta,meta=Label, col sep=comma]{data/time_shift.csv};

            \end{axis}
\end{tikzpicture}
 }
 \caption{Time shifting of local time stamp counter.}
 \label{fig:timeshift}
\end{figure}
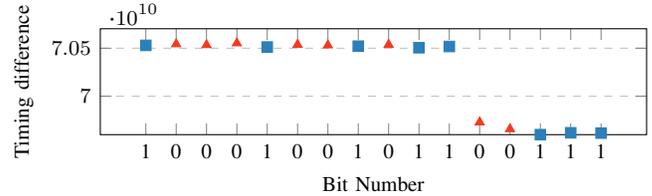

We first evaluate the amplification in JavaScript directly in the V8 engine.
We use an amplification factor of \SIx{250000} and use a native timestamp counter to measure the response times.
We run a script for a uniformly distributed secret with \SI{16}{\bit} length.
For the first 11 bits, we observe an average timing difference of \SI{21779307}{cycles}.
However, we observe that the execution timings of the function shift over time, even on an isolated core with a fixed CPU frequency (\cf
\cref{fig:timeshift} \Cref{sec:appendix:timeshift}). 
We measure the runtime of our amplified JavaScript attack and estimate the maximum achievable leakage rate for our script.
All evaluated numbers are provided in~\cref{tab:js_attack} (\Cref{sec:appendix:limits}).
One script execution takes about \SI{30}{\second}, and with a success rate of \SI{100}{\percent} we determine an optimal leakage rate of \SI{2}{\bit/\minute} leading to a leakage rate of \SI{120}{\bit/\hour}.

The current default settings for \CFWorkers are provided in~\cref{tab:current_limits} (\Cref{sec:appendix:limits}).
With these settings, we cannot mount a successful attack due to the short runtime (\SI{50}{\milli\second}).
However, \CFWorkers consider higher runtimes and a larger amount of subrequests per intercepted request in the future. %
We leave it as future work to additionally improve the attack performance by leveraging parallelism in the attack~\cite{VanGoethem2020Timeless}.
Generally, our attack is not limited to \CFWorkers. 
It shows that an attacker does not need access to a local timer to mount microarchitectural attacks. 
Hence, this complements the results from Schwarz~\etal\cite{Schwarz2019netspectre} showing that an attacker does not require code execution, and McIlroy~\etal\cite{Mcilroy2019} showing that microarchitectural timings can be amplified and exploited with a low-resolution timer.

\section{\DefenseName}\label{sec:isolation}
In this section, we present an approach to dynamically isolate malicious \CFWorkers to benefit both from the security of process isolation and the performance of language-level isolation. 
The basic idea is to use HPCs to detect potential Spectre attacks and isolate suspicious \CFWorkers using process isolation.
While a detection mechanism typically suffers from false positives, \DefenseName can cope even with high false-positive rates.
In the worst case, a Spectre attack is detected for every worker, leading to the worst-case scenario of one \worker per process, \ie strict process isolation, as currently also used in browsers plus the \SI{2}{\percent} detection overhead.
As \workers are stateless, they can also be suspended or migrated at any time.
Thus, even if many \worker are considered malicious, the resources of \CF are not exhausted. 
If the false positive rate is below \SI{100}{\percent}, the overall performance is better than for strict process isolation.

First, we discuss how to reliably detect Spectre attacks using performance counters (\cf \Cref{sec:dwi:detect}).
Second, we perform a microbenchmark on the \texttt{perf} interface to evaluate its overhead on different CPUs with different kernel boot parameters and microcode versions (\cf \Cref{sec:dwi:overhead}).
Finally, we integrate our approach into \CFWorkers and measure the performance overhead of reading out performance counters on a real-world cloud system (\cf \Cref{sec:dwi:perf}).
We show that there is a negligible performance overhead of \SI{2}{\percent} for reading out performance counters.

\begin{figure}[t]
 \centering
 \resizebox{\hsize}{!}{
    \input{images/adaptive_site_isolation.tikz}
}
\caption{\DefenseName isolating a malicious \worker based on the performance-counter reading.}
\label{fig:adaptive_site_isolation}
\end{figure}

\subsection{Detecting Spectre Attacks}\label{sec:dwi:detect}
In this section, we discuss the detection of Spectre attacks using HPCs.
While the common use of HPCs is finding bottlenecks, researchers used HPCs for detecting malware, rootkits, CFI violations, ROP, Rowhammer, or cache-side channel attacks~\cite{Wang2013Numchecker,Xia2012,Malone2011,Zhou2014HDROP,Herath2015,Briongos2018CacheShield,Gruss2016Flush,Chiappetta2015}.
For \DefenseName, we develop two approaches based on HPCs.
The first approach (\cf \Cref{sec:detect:histogram}) samples the mispredicted and retired branches and compares them to the distribution of a template attack using a two-sided Kolmogorov-Smirnov (KS) test~\cite{Richard2011TwoSidedKS}.
Our second, faster approach uses the number of retired branches normalized with the iTLB accesses, similarly to the approach used by Gruss~\etal\cite{Gruss2016Flush} to detect cache attacks.

\subsubsection{Detecting Attacks using PEBS}\label{sec:detect:histogram}
\begin{sloppypar}
Our first approach uses Precise Event-Based Sampling (PEBS) on all retired and mispredicted branches (\texttt{BR\_MISP\_RETIRED.ALL\_BRANCHES\_PEBS}).
Based on the samples acquired using PEBS, we build a simple pattern detection for Spectre-PHT gadgets with in-place mistraining~\cite{Canella2019A}.
In such a scenario, an attacker first executes the bounds check multiple times with an in-bounds index to mistrain the branch predictor.
Then, for the actual transient out-of-bounds access, the branch predictor predicts that the index is in-bounds with a high probability.
We only focus on the in-place variant of Spectre-PHT, as it is the most stable variant that cannot be mitigated in microcode.
\end{sloppypar}

\begin{figure}[t]
    \centering
    \resizebox{0.8\hsize}{!}{
    \resizebox{\hsize}{!}{
\begin{tikzpicture}
            \begin{axis}[
            ybar,
            bar width=0.5cm,
            width={\hsize},
            height=3cm,
            xlabel={Branch ID},
            ylabel={Mispredictions},
            restrict x to domain=-1:10.5,
            xmin=0,
            ymin=0,
            xmax=10.5,
            legend pos=north west,
            legend style={at={(0.85,0.95)}}
            ]
            \addplot[thick,color=blue] table[x index = {0}, y index = {1}, col sep=comma]{data/nero2d.csv};
            \end{axis}
\end{tikzpicture}
}
    }
    \caption{Distribution of mispredicted and retired branches for the Nero2D program~\cite{Phoronix2020Benchmark}.}
    \label{fig:distribution_nero}
\end{figure}
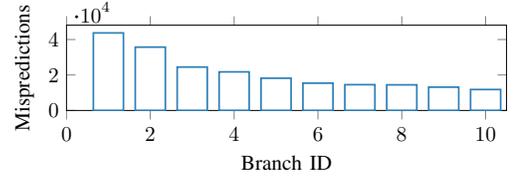

{\sloppy
We use \texttt{BR\_MISP\_RETIRED.ALL\_BRANCHES\_PEBS}  (\texttt{EventSel=C5H}, \texttt{UMask=04H}) as a counter, and set the event count to 1 to sample every mispredicted and retired branch.
Each sample contains the virtual address of a branch that was initially mispredicted but retired later on.
Hence, this approach provides the per-branch information of the number of mispredictions.
}
\cref{fig:distribution} illustrates the histogram of mispredicted and retired branches of a Spectre gadget running a Spectre-PHT attack for a minute.
As shown in~\cref{fig:distribution_nero}, there is a strong difference in the distribution of the mispredicted branches for a benign program.
We use the two-sided KS-test for discrete distributions to compare the distribution of a sample Spectre attack and sample programs.
This test verifies the hypothesis of whether two independent samples are drawn from the same independent distribution.
To reject the hypothesis that the two distributions are equal, the $p$-value can be chosen to a certain value.
The smaller the $p$-value, the higher the confidence, typically it is chosen to be between $0.01$ and $0.1$ to achieve a higher confidence.
We observe that a typical Spectre attack only has a few highly mispredicted branches.
Contrary to our expectations, the most mispredicted branch is not the branch exploited by the Spectre attack, \ie the bounds check.
Instead, it is the delay loop used to increase the transient window (\cf Appendix C, line 55 in Kocher~\etal\cite{Kocher2019}).
On the second rank is the branch that is mistrained.
This delay loop or similar functions are required and important for the success rate of the Spectre attack.
If a delay loop which counts up to 100 is used, the success rate of the Spectre attack is at \SI{99.76}{\percent} ($n=1000$, $\sigma_{\bar{x}}$=\SI{0.53}{\percent}).
If omitted, the success rate of the Spectre goes down to nearly zero (\SI{0.64}{\percent}, $n=1000$, $\sigma_{\bar{x}}$ = \SI{1.47}{\percent}).
Moreover, the history of recently taken/non-taken branches affects branch prediction. 
Hence, adding a fixed-size delay loop helps controlling the branch-predictor  state, ensuring consistent conditions for the mistrained branch~\cite{Tsuro2021SpectreJS}.
Thus, also in JavaScript, this delay loop is required for a successful attack.
We evaluate this approach in~\cref{sec:eval_histogram}.

\begin{figure}[t]
    \centering
    \resizebox{0.8\hsize}{!}{
    \resizebox{\hsize}{!}{
\begin{tikzpicture}
            \begin{axis}[
            ybar,
            bar width=0.5cm,
            width={\hsize},
            height=3cm,
            xlabel={Branch ID},
            ylabel={Mispredictions},
            xmin=0,
            ymin=0,
            xmax=10,
            legend pos=north west,
            legend style={at={(0.85,0.95)}}
            ]
            \addplot[thick,color=blue] table[x index = {0}, y index = {1}, col sep=comma]{data/spectre_distrib.csv};
            \end{axis}
\end{tikzpicture}
}
    }
    \caption{Distribution of mispredicted and retired branches in a Spectre-PHT attack sorted by the number of mispredictions. The branch ID uniquely identifies a branch at a specific virtual address and is assigned based on the number of mispredictions, \ie, branch ID 1 refers to the branch with the highest number of mispredictions.}
    \label{fig:distribution}
\end{figure}
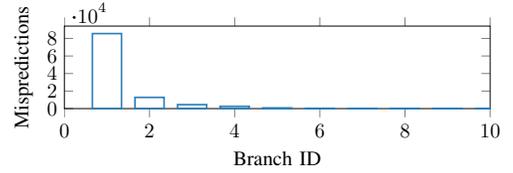

\subsubsection{Detecting Attacks using Normalized Performance Counters}
\begin{sloppypar}
Our second approach tries to detect Spectre attacks using normalized performance counters.
At first we collect data from different performance counters.
{\sloppy
We collect the following hardware events (\texttt{PERF\_COUNT\_HW\_*}): \texttt{CACHE\_iTLB}, \texttt{BRANCH\_MISSES},  \texttt{BRANCH\_INSTRUCTIONS}, \texttt{CACHE\_REFERENCES}, \texttt{CACHE\_MISSES}, \texttt{CACHE\_L1D/READ\_MISSES} and \texttt{CACHE\_L1D/READ\_ACCESSES}.}
We normalize the values using iTLB performance counters (iTLB accesses) which was also used by Gruss~\etal\cite{Gruss2016Flush} to detect Rowhammer and cache attacks.
Similarly to Rowhammer and cache attacks, the main attack code for Spectre has a small code footprint with a high activity in the branch-prediction unit.
\end{sloppypar}

The iTLB counter normalizes the branch-prediction events with respect to the code size by dividing the performance counter value by the number of iTLB accesses. %
We integrate the monitor into \CFWorkers, to read the performance counters before and after each script execution. 
The averaged per-execution numbers are updated in a 1-second interval (Note that a single script runs per default \SI{50}{\milli\second}, \cf ~\cref{tab:current_limits} \cref{sec:appendix:limits}).
While reducing the interval does not directly impact the performance of a \worker, it potentially leads to more false positives as outliers are not filtered. 
We collect data from the benign workload and compare it to a \worker executing a Spectre attack.
Based on the performance numbers, we find a threshold to distinguish between an attack and normal workload.
We evaluate this approach in~\cref{sec:eval_normalized_ctr} and discuss further detection methods in~\cref{sec:other_methodologies}.

\subsection{Overhead of Hardware Performance Counters}\label{sec:dwi:overhead}

In this section, we evaluate the real-world performance overhead of using performance counters as a detection mechanism.
Most previous works use smaller microbenchmarks or smaller real-world applications~\cite{Zhang2013,Irazoqui2018mascat,Payer2016,Zhang2016CloudRadar,Chiappetta2015,Herath2015,Mushtaq2020}.
In addition to microbenchmarks, we also evaluate the performance on \CFWorkers to show that the overheads of many performance counters are so large that they cannot be used in a real-world cloud system.

\paragrabf{Setup.}
We run a C microbenchmark~\cite{CBenchmarksGame} computing points on a Mandelbrot set. %
Such a CPU-bound benchmark is similar to a \worker workload, as they cannot directly issue syscalls or interact with I/O devices. 
First, we set up the performance counters and measure the overhead of the setup.
In a second test, we activate one performance counter and read out the values. %
Most Intel architectures are limited to program 4 HPCs per core, or 8 HPCs per core if hyperthreading is disabled~\cite{Intel2017Performance}.
Therefore, we activate up to four different counters and measure their performance overhead.
In addition, we run our programs multi-threaded to see the additional overhead.
We use \texttt{PERF\_SAMPLE\_READ} as sample type for our test cases such that all counters in a group are read.
We repeat the microbenchmark $100$ times and measure the average execution time.
{\sloppy
Additionally, we evaluate the impact of different transient-execution-attack countermeasures on the overhead of performance counters.
We evaluate $5$ different microcode versions on Intel CPUs that introduced mitigations, ranging from January 2018 to April 2020.}
In addition, we compare the impact of different active transient-execution-attack mitigations supported by the Linux kernel by successively disabling them.\footnote{\scriptsize\url{https://www.kernel.org/doc/html/latest/admin-guide/kernel-parameters.html}} %
We run our benchmark on a Xeon E5-1630.

\paragrabf{Experimental Results.}
To mitigate the Rogue System Register Read vulnerability (CVE-2018-3640), a microcode update ensures that \texttt{rdmsr} cannot be used transiently~\cite{IntelRSRRadvisory}.
Hence, we expect a measurable difference caused by the microcode when reading performance counters using \texttt{rdmsr}.
In a microbenchmark, we measure the time it takes to read a performance counter \SIx{1000000} times, once using \texttt{rdmsr} from user space via the \texttt{msr} module, and once using the user-space-enabled \texttt{rdpmc} instruction.
For the \texttt{rdmsr} benchmark, we measure the time to perform the syscall \texttt{pread} whichs reads out the msr value.
Before the patched microcode (Ngovember 2017), the reading takes 37 cycles for \texttt{rdpmc} and 1291 cycles for \texttt{rdmsr}.
Note that \texttt{rdmsr} is handled by the kernel, hence the larger overhead compared to \texttt{rdpmc}.
Since the microcode update containing the mitigation (October 2019 update), the reading takes on average 37 cycles for \texttt{rdpmc} and 2003 cycles with \texttt{rdmsr}.
Hence, the mitigation has indeed slowed down the \texttt{rdmsr} instruction, while it did not affect \texttt{rdpmc}.
As all other mitigations are disabled during this test, the overhead can be attributed to the changed microcode.
However, for the microbenchmark where we sample the performance counters only once every \SI{50}{\milli\second}, the performance overhead is amortized for all microcode versions.
The overhead of reading the performance counters before and after the benchmark compared to normally running the benchmark was on average \SI{0.04}{\percent} ($n = 100$, $\sigma_{\bar{x}} =$ \SI{0.011}{\percent}).

\begin{figure}[t]
 \centering
 \resizebox{\hsize}{!}{
    \resizebox{\hsize}{!}{
\begin{tikzpicture}[yscale=1]
\pgfplotstableread[col sep=comma,]{data/resultCounterSampleXeon.csv}\datatable
 \begin{axis}[
     ybar,
     height = 4cm,
     width = 0.7\textwidth,
     ymin = 0.9998,
     ystep=0.001,
     ymax = 1.001,
   xtick=data,
   legend pos=north west,
     xticklabels from table={\datatable}{Countermeasures},
     x tick label style={font=\normalsize, rotate=30, anchor=east},
     yticklabel=\pgfmathparse{100*\tick}\pgfmathprintnumber{\pgfmathresult}\,\%,
     yticklabel style={/pgf/number format/.cd,fixed,precision=2}]
    ]
      \addplot[fill=gray] table [x expr=\coordindex, y={Xeon E5-1630}]{\datatable};
      \addplot[fill=white] table [x expr=\coordindex, y={Xeon E5-1630-dis}]{\datatable};
      \addlegendentry{Activated counters}
    \addlegendentry{Baseline}
  \end{axis}
\path (current bounding box.north west)
      -- node[yshift=3ex,xshift=-2ex,rotate=90] {Average Runtime}
      (current bounding box.south west);
\end{tikzpicture}
}
 }
\caption{HPC microbenchmark with different mitigations. The grey bars show the scaled performance overhead in comparison to the baseline programs (white bars).}
\label{fig:mitigations}
\end{figure}
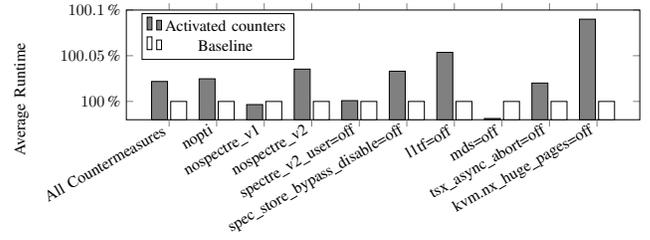

In our second experiment, we selectively disable kernel mitigations for transient-execution attacks and evaluate their impact on the performance of reading performance counters.
We use the microcode version from January 2020.
We observe an additional average overhead of \SI{0.08}{\percent} when running with all Spectre mitigations enabled, as shown in \cref{fig:mitigations}.
When reading the same performance counters directly in the kernel, we observe a small averaged performance overhead of \SI{2.2}{\percent} ($n=1000$, $\sigma_{\bar{x}}=0.5\,\%$) on our test devices.

We integrate the performance counter setup into \CFWorkers and run a JavaScript benchmark to evaluate its performance in a real-world cloud scenario.
Reading the performance-counter values from the \texttt{perf} subsystem results in a high performance overhead of \SI{22}{\percent} ($n=$\SIx{10000}).
If the counter is additionally enabled and disabled before and after the script execution, respectively, the overhead increases to \SI{33}{\percent} ($n=$\SIx{10000}).
Hence, using the \texttt{perf} subsystem for performance-counter-based detection mechanisms is too expensive for \CFWorkers.
With always-enabled counters and \texttt{rdpmc}, we measure a performance overhead of \SI{2}{\percent} ($n=$\SIx{10000}).
Thus, low-overhead detection with HPCs for a production system requires obtaining measurements directly from the kernel or the user-space-enabled \texttt{rdpmc} instruction.

\subsection{Process Isolation}\label{sec:dwi:perf}

For \DefenseName, we fundamentally rely on process isolation.
A well-known implementation of process isolation is site isolation, where every page in a browser runs in its own process to prevent memory safety violations as well as Spectre attacks~\cite{Reis2018siteisolation}.
However, in contrast to full site isolation, we only isolate malicious \CFWorkers if the Spectre detection mechanism flags them.
Hence, Dynamic Process Isolation only falls back to full site isolation in the worst case, while reducing the overhead caused by process isolation in the average case where only suspicious \CFWorkers are isolated.

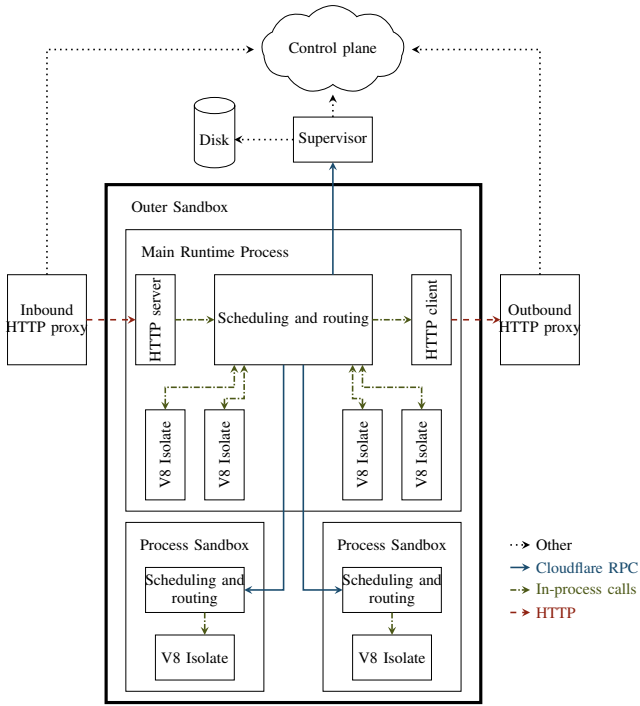
\begin{figure}
\resizebox{\hsize}{!}{
 \resizebox{!}{\vsize}{
\begin{tikzpicture}[yscale=0.8,xscale=0.7]
\footnotesize
\usetikzlibrary{shapes}

\draw (0,0) rectangle +(4,2) node [midway] {Scheduling and routing};
\draw (5,0) rectangle +(1,2) node[midway,rotate=90] {HTTP client};
\draw (-2,0) rectangle +(1,2) node [midway,rotate=90] {HTTP server};

\draw (-5.25,0) rectangle +(2,2) node[midway] {\parbox{1.5cm}{\centering Inbound HTTP proxy}};
\draw (7.25,0) rectangle +(2,2) node[midway] {\parbox{1.5cm}{\centering Outbound HTTP proxy}};

\draw (-1.75,-3) rectangle +(1,2) node[midway,rotate=90] {V8 Isolate};
\draw (-0.25,-3) rectangle +(1,2) node[midway,rotate=90] {V8 Isolate};
\draw (4.75,-3) rectangle +(1,2) node[midway,rotate=90] {V8 Isolate};
\draw (3.25,-3) rectangle +(1,2) node[midway,rotate=90] {V8 Isolate};

\draw (-2.25,-3.25) rectangle +(8.5,6.25);
\node[right] at (-2,2.5) {Main Runtime Process};

\draw (-2.25,-7.25) rectangle +(3.5,3.75);
\draw (-1.75,-5.5) rectangle +(2.5,1) node[midway] {\parbox{2cm}{\centering Scheduling and routing}};
\draw (-1.5,-7) rectangle +(2,1) node[midway] {V8 Isolate};

\node at (-0.5,-4) {\parbox{3cm}{\centering Process Sandbox}};

\draw (2.75,-7.25) rectangle +(3.5,3.75);
\draw (3.25,-5.5) rectangle +(2.5,1) node[midway] {\parbox{2cm}{\centering Scheduling and routing}};
\draw (3.5,-7) rectangle +(2,1) node[midway] {V8 Isolate};

\node at (4.5,-4) {\parbox{3cm}{\centering Process Sandbox}};

\draw[ultra thick] (-2.75,-7.5) rectangle +(9.5,11.5);
\node[right] at (-2.25,3.5) {Outer Sandbox};

\draw (2,4.5) rectangle +(2,1) node[midway] {Supervisor};

\node[cloud,draw,cloud puffs=9,cloud puff arc=100, aspect=2] at (3,7) {Control plane};
\node[cylinder,draw,shape border rotate=90,aspect=0.4,minimum height=1.25cm] at (0,5) {Disk};

\draw[thick,densely dashdotted,green!40!black,->,>=stealth] (-1,1) to (0,1);
\draw[thick,densely dashdotted,green!40!black,->,>=stealth] (4,1) to (5,1);
\draw[thick,dashed,red!60!black,->,>=stealth] (6,1) to (7.25,1);
\draw[thick,dashed,red!60!black,->,>=stealth] (-3.25,1) to (-2,1);
\draw[thick,densely dashdotted,green!40!black,<->,>=stealth] (-1.25,-1) |- (0.5,-0.5) -- (0.5,0);
\draw[thick,densely dashdotted,green!40!black,<->,>=stealth] (0.25,-1) |- (0.75,-0.75) -- (0.75,0);
\draw[thick,densely dashdotted,green!40!black,<->,>=stealth] (3.75,-1) |- (3.5,-0.75) -- (3.5,0);
\draw[thick,densely dashdotted,green!40!black,<->,>=stealth] (5.25,-1) |- (3.75,-0.5) -- (3.75,0);

\draw[thick,->,blue!60!black,>=stealth] (1.75,0) |- (0.75,-5);
\draw[thick,->,blue!60!black,>=stealth] (2.25,0) |- (3.25,-5);

\draw[thick,densely dashdotted,green!40!black,->,>=stealth] (-0.25,-5.5) -- (-0.25,-6);
\draw[thick,densely dashdotted,green!40!black,->,>=stealth] (4.5,-5.5) -- (4.5,-6);

\draw[thick,dotted,->,>=stealth] (8.25,2) |- (5,7);
\draw[thick,dotted,->,>=stealth] (-4.25,2) |- (1,7);
\draw[thick,->,blue!60!black,>=stealth] (3,2) -- (3,4.5);
\draw[thick,dotted,->,>=stealth] (3,5.5) -- (3,6);
\draw[thick,dotted,->,>=stealth] (2,5) -- (0.5,5);

\draw[thick,dotted,->,>=stealth] (7.5,-4) -- (8,-4) node[right] {Other};
\draw[thick,->,blue!60!black,>=stealth] (7.5,-4.5) -- (8,-4.5) node[right] {Cloudflare RPC};
\draw[thick,densely dashdotted,green!40!black,->,>=stealth] (7.5,-5) -- (8,-5) node[right] {In-process calls};
\draw[thick,dashed,red!60!black,->,>=stealth] (7.5,-5.5) -- (8,-5.5) node[right] {HTTP};

\end{tikzpicture}
}
 }
 \caption{System overview for \DefenseName.}
 \label{fig:overview}
\end{figure}

Related work proposes efficient in-process isolation mechanisms using Intel Memory Protection Keys (MPK)~\cite{VahldiekOberwagner2019,Park2018libmpk,Schrammel2020Donky}.
However, Intel MPK is only available on selected CPUs since Skylake-SP, limited to 16 protection keys and thus not practical for \CFWorkers~\cite{VahldiekOberwagner2019}, running multiple thousand \workers per process.
Furthermore, the threat model of these approaches does not include side-channel or transient-execution attacks.
To perform dynamic process isolation, we modify the \CFWorkers software to isolate a potentially malicious \worker, \ie a \worker that was flagged by the performance-counter-based detection, into a separate process.
We implement process isolation in \CFWorkers from scratch, as illustrated in \Cref{fig:overview}.
For that, we start process sandboxes by forking from a zygote process, and talk to the new process over an RPC protocol~\cite{Varda2019CapnProtoAnon,Cloudflare2020AnonProcessIsolationBlog}.
All communication between the main process and the isolated process are over this RPC connection, communications between the process sandbox and the outside world have to go back through the main process.
Since the runtime of a \worker is, on average, less than \SI{1}{\milli\second}, the isolation must not introduce a high performance overhead.
Thus, one instance of a \worker frequently reads out the performance counters per script execution and calculates a moving average.
From our results in \cref{sec:evaluation}, we observed that the normalized iTLB performance provides the best detection tradeoff in terms of performance overhead and accuracy.
We first run an attack and collect its performance-counter data.
Additionally, we collect the per-CPU-core performance-counter data of real scripts running on the production system.
Based on our evaluation in \cref{sec:eval_normalized_ctr}, we use a threshold of 4096 retired branches per iTLB access to distinguish between a suspicious and a benign script.

If a script exceeds this threshold, we flag it as a potential Spectre attack.
Such a \worker is isolated into a separate process.
In contrast to, \eg browser tabs, \worker are stateless. 
Thus, a \worker can simply be migrated. 
Isolating instead of terminating ensures that the \worker can still continue running, \eg in case the detection was a false positive, while it cannot access data of any other \worker.

\section{Evaluation} \label{sec:evaluation}

In this section, we evaluate the accuracy and performance overhead of our detection methodologies.
First, we evaluate our PEBS-based approach (\cf \Cref{sec:detect:histogram}).
We choose a threshold of \SIx{4096} branch accesses per iTLB access, which allows distinguishing a Spectre attack from a benign script.
We use a large set of different programs to sample the number of mispredicted and retired branches.
For our set, we observe that out of \SIx{141} programs, which includes the 13 Spectre gadgets from Kocher~\cite{Kocher2018mitigations}, we cannot distinguish \SIx{4} benign programs from a Spectre gadget, resulting in a false-positive rate of \SI{2.83}{\percent} with a small performance overhead of \SI{2}{\percent}.
However, in our production environment, we observe a performance overhead of up to \SI{50}{\percent}.
In contrast, for our normalized counters approach, we observe a negligible overhead of \SI{2}{\percent}. %

\subsection{PEBS-based Approach} \label{sec:eval_histogram}
We evaluate the PEBS-based approach using \SIx{128} programs from the Phoronix benchmark suite~\cite{Phoronix2020Benchmark} (\cf \cref{tab:evaluated_programs} \Cref{sec:appendix:phoronix}).
The test suite already supports the perf interface. 
We only have to add the \texttt{BR\_MISP\_RETIRED.ALL\_BRANCHES\_PEBS} event.

We use the two-sided KS-test for discrete distributions~\cite{Richard2011TwoSidedKS} to compare the distribution of mispredictions.
In our scenario, we compare the distribution of mispredictions for benign and malicious scripts.
We compare the programs from \cref{tab:evaluated_programs} (\Cref{sec:appendix:phoronix}) to our Spectre attack.
In addition, we integrate $13$ sample Spectre gadgets from Kocher~\cite{Kocher2018mitigations} and use it to evaluate the detection.\footnote{Only $13$ out of $15$ test gadgets leaked data on our evaluated system.}

We use the top $100$ branches of the programs as input to the KS-statistic.
If the $p$-value is high or the KS-statistic is low, we cannot reject the hypothesis that the two distributions are the same.
We choose the $p$-value to be $0.1$ to achieve a \SI{90}{\percent} confidence level.
Every script with a $p$-value lower than this threshold is rejected.
The $p$-value seems to be high, however, as the structure of certain test programs have certain similarities to a Spectre attack, this filters out most of the programs with a small similarity.
Thus, if the $p$-value is below this threshold, we can reject the hypothesis that the two samples are drawn from the same distribution.
Conversely, if the $p$-value is above the threshold, we cannot reject the hypothesis and thus count it as a false positive.
We evaluate all $141$ test programs using the KS-test.
We detect all $13$ test programs and achieve a true positive rate of \SI{100}{\percent}.
Using this approach, we achieve a false positive rate of \SI{2.83}{\percent}.
In total, we achieve an F-Score of \SIx{0.87}.
We integrate this approach into \CFWorkers system.
We sample the PEBS events on each occurrence while the script is running to get an exact distribution.
To evaluate the performance overhead, we first run this sampling approach on a single-threaded Spectre-PHT attack proof of concept.
For the small proof of concept, we measure a performance overhead of \SI{2.3}{\percent} ($n$=1000, $\sigma_{\bar{x}}$=\SI{0.5}{\percent})
However, on a production system, we get an overhead of \SI{50}{\percent} ($n$=100, $\sigma_{\bar{x}}$=\SI{8.13}{\percent}).
While the approach works well in smaller non-multithreaded environments, an overhead of \SI{50}{\percent} is too costly to run it on the production environment.

Copying the data in batches or directly mapping the PEBS buffer into user space\footnote{\scriptsize\url{https://elixir.bootlin.com/linux/v5.8/source/arch/x86/events/intel/ds.c#L1754}} could drastically improve the performance of PEBS sampling, decreasing the performance overhead of this approach.
However, we did not evaluate this on the \CFWorkers production system, as modifying the kernel on a production system would be a huge engineering and testing effort to ensure that no customer data is at risk.

\subsection{Normalized Performance Counters} \label{sec:eval_normalized_ctr}
The PEBS-based approach is too expensive on a real-world cloud system.
However, as shown in~\cref{sec:evaluation}, the performance overhead is only \SI{2}{\percent} when using the \texttt{rdpmc} instruction for reading raw performance-counter values.
We verified this on $5$ Intel Xeon server CPUs (Broadwell, Skylake 4116, Skylake 6162, Skylake 6162, Cascade Lake 6262) and one AMD Epyc Rome CPU running in the cloud.
To decide whether a script is susceptible or not, we collect performance data from the production system running our Spectre attack.
We recorded the performance counters on the production environment and sampled over \SIx{50000} times as a baseline.

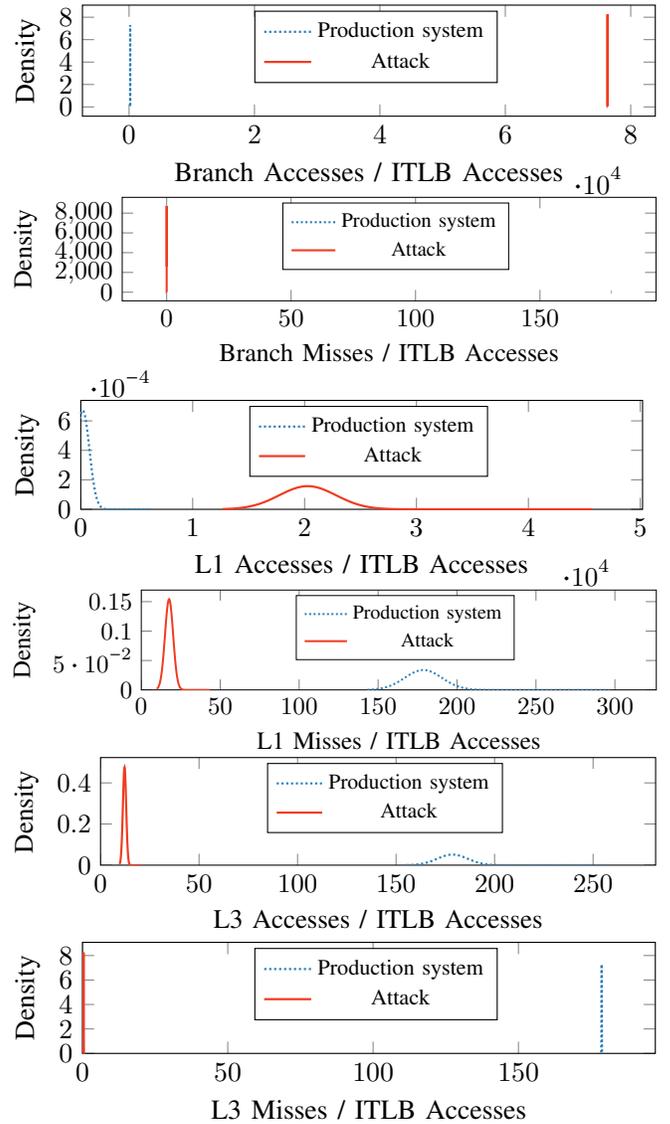
\begin{figure}[t]
 \centering
    \begin{subfigure}{\hsize}%
        \resizebox{\hsize}{!}{
\begin{tikzpicture}
            \begin{axis}[
            width={\hsize},
            height=3cm,
            xlabel={Branch Accesses / ITLB Accesses},
            ylabel={Density},
            legend style={at={(0.30,0.95)},anchor=north west,font=\footnotesize}
            ]
            \addplot[thick,color=blue,densely dotted] table[x index = {0}, y index = {1}, col sep=comma]{data/btb_accesses_cf.txt};
            \addplot[thick,color=red]  table[x index = {0}, y index = {1}, col sep=comma]{data/btb_accesses_attack.txt};

            \legend{Production system,Attack}
            \end{axis}
\end{tikzpicture}
}%
    \end{subfigure}
    \begin{subfigure}{\hsize}%
        \resizebox{\hsize}{!}{
\begin{tikzpicture}
            \begin{axis}[
            width={\hsize},
            height=3cm,
            xlabel={Branch Misses / ITLB Accesses},
            ylabel={Density},
            legend style={at={(0.30,0.95)},anchor=north west,font=\footnotesize}
            ]
            \addplot[thick,color=blue,densely dotted] table[x index = {0}, y index = {1}, col sep=comma]{data/btb_misses_cf.txt};
            \addplot[thick,color=red]  table[x index = {0}, y index = {1}, col sep=comma]{data/btb_misses_attack.txt};

            \legend{Production system,Attack}
            \end{axis}
\end{tikzpicture}
}%
    \end{subfigure}
    \begin{subfigure}{\hsize}%
        \resizebox{\hsize}{!}{
\begin{tikzpicture}
            \begin{axis}[
            width={\hsize},
            height=3cm,
            xlabel={L1 Accesses / ITLB Accesses},
            ylabel={Density},
            xmin=0,
            ymin=0,
            legend style={at={(0.30,0.95)},anchor=north west,font=\footnotesize}
            ]
            \addplot[thick,color=blue,densely dotted] table[x index = {0}, y index = {1}, col sep=comma]{data/l1_accesses_cf.txt};
            \addplot[thick,color=red]  table[x index = {0}, y index = {1}, col sep=comma]{data/l1_accesses_attack.txt};

            \legend{Production system,Attack}
            \end{axis}
\end{tikzpicture}
}%
    \end{subfigure}
    \begin{subfigure}{\hsize}%
        \resizebox{\hsize}{!}{
\begin{tikzpicture}
            \begin{axis}[
            width={\hsize},
            height=3cm,
            xlabel={L1 Misses / ITLB Accesses},
            ylabel={Density},
            xmin=0,
            ymin=0,
            legend style={at={(0.30,0.95)},anchor=north west,font=\footnotesize}
            ]
            \addplot[thick,color=blue,densely dotted] table[x index = {0}, y index = {1}, col sep=comma]{data/l1_misses_cf.txt};
            \addplot[thick,color=red]  table[x index = {0}, y index = {1}, col sep=comma]{data/l1_misses_attack.txt};

            \legend{Production system, Attack}
            \end{axis}
\end{tikzpicture}
}%
    \end{subfigure}
    \begin{subfigure}{\hsize}%
        \resizebox{\hsize}{!}{
\begin{tikzpicture}
            \begin{axis}[
            width={\hsize},
            height=3cm,
            xlabel={L3 Accesses / ITLB Accesses},
            ylabel={Density},
            xmin=0,
            ymin=0,
            legend style={at={(0.30,0.95)},anchor=north west,font=\footnotesize}
            ]
            \addplot[thick,color=blue,densely dotted] table[x index = {0}, y index = {1}, col sep=comma]{data/l3_access_cf.txt};
            \addplot[thick,color=red]  table[x index = {0}, y index = {1}, col sep=comma]{data/l3_access_attack.txt};

            \legend{Production system,Attack}
            \end{axis}
\end{tikzpicture}
}%
    \end{subfigure}
    \begin{subfigure}{\hsize}%
        \resizebox{\hsize}{!}{
\begin{tikzpicture}
            \begin{axis}[
            width={\hsize},
            height=3cm,
            xlabel={L3 Misses / ITLB Accesses},
            ylabel={Density},
            xmin=0,
            ymin=0,
            legend style={at={(0.30,0.95)},anchor=north west,font=\footnotesize}
            ]
            \addplot[thick,color=blue,densely dotted] table[x index = {0}, y index = {1}, col sep=comma]{data/l3_misses_cf.txt};
            \addplot[thick,color=red]  table[x index = {0}, y index = {1}, col sep=comma]{data/l3_misses_attack.txt};
            \legend{Production system, Attack}
            \end{axis}
\end{tikzpicture}
}%
    \end{subfigure}
\caption{Performance counters of the average \CFWorkers script and a Spectre attack script on the production system.}
\label{fig:leaked_byte}
\end{figure}

\cref{fig:leaked_byte} shows the normalized performance-counter values of our cloud machines.
In the case of last-level-cache accesses, last-level-cache misses, and branch misses, the numbers of the attack script are below the average script.
For the number of L1-cache accesses and retired branches, we can clearly distinguish between an average script and the attack.
Especially for the retired branches, the distance between an attack script and the average regular script is \SI{34} times the standard deviation of a benign script.

We choose the number of normalized retired branch instructions as an indicator for a Spectre attack and run it on our cloud machines.
First, we run a Spectre attack to verify whether their number is in a similar range on each test machine.
We then evaluate different threshold boundaries for the number of normalized retired branch instructions and report the number of false positives.

\begin{figure}[t]
    \centering
    \resizebox{\hsize}{!}{
    \resizebox{\hsize}{!}{
\begin{tikzpicture}[transform shape]
\begin{axis}[
    xlabel={Threshold},
    width  = {\hsize},
    height = 4cm,
    ylabel={False Positives [\%]},
    legend style={font=\tiny},
    legend pos=north east,
    ymajorgrids=true,
    grid style=dashed,
    scaled x ticks = false,
    x tick label style={font=\normalsize, rotate=30, anchor=east,yshift=-0.3em}
]
\addplot[red,mark=x] coordinates {(1024,4.671) (2048,0.6) (4096,0.453) (8192,0.453)}; %
\addplot[green,mark=*] coordinates {(1024,7.936) (2048,4.29) (4096,0.7) (8192,0.16)};
\addplot[blue,mark=diamond] coordinates {(1024,25.9) (2048,22.85) (4096,0.85) (8192,0.28)};
\addplot[orange,mark=o] coordinates {(1024,26.14) (2048,23.86) (4096,0.57) (8192,0.21)};
\addplot[black,mark=triangle] coordinates {(1024,24.05) (2048,21.94) (4096,0.7) (8192,0.2)};
\addplot[brown,mark=diamond*] coordinates {(1024,21.63) (2048,0.81) (4096,0.29) (8192,0.22)};

\legend{Broadwell (7),Skylake 4116 (8),Skylake 6162 (9 lbg-1g),Skylake 6162 (9 lbg-4),Cascade Lake(9.5),AMD Epyc Rome (10)}
\end{axis}
\end{tikzpicture}
}
    }
    \caption{Number of false positives depending on the normalized iTLB threshold.}
    \label{fig:fp}
\end{figure}
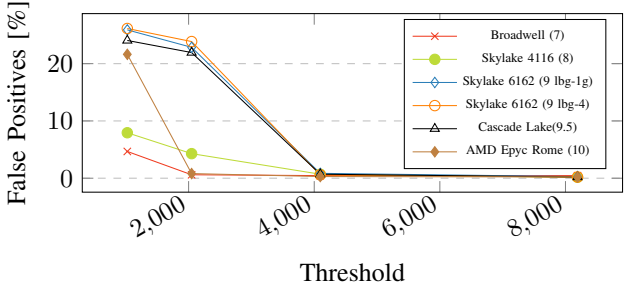

\cref{fig:fp} shows the number of false positives depending on the threshold on our cloud machines in the production environment.
For a strict threshold, \ie 1024, the false-positive rate is \SI{21.41}{\percent}.
However, this threshold can be set higher to reduce the number of false positives.
The numbers of false positives are in a similar range on each of the tested machines. %
Setting the threshold to \SIx{4096}, results in an average false positive rate of \SI{0.61}{\percent} on our devices.
For a threshold of \SI{8192} the average false-positive rate decreases to \SI{0.26}{\percent}, and at a threshold of \SI{65536}, we do not observe any false positives.

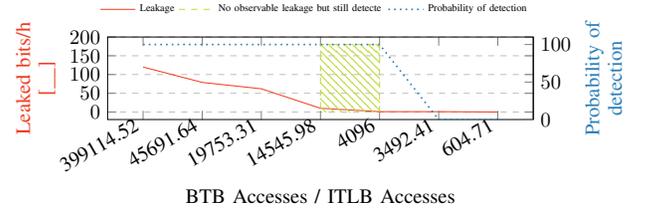
\begin{figure}[t]
    \centering
    \resizebox{\hsize}{!}{
    \pgfplotsset{compat=newest}
\resizebox{\hsize}{!}{
\begin{tikzpicture}[transform shape]
\pgfplotsset{
    legend style={draw=none,legend cell align=left,}
}
\begin{axis}[
    xlabel={BTB Accesses / ITLB Accesses},
    width  = {\hsize},
    height = 3cm,
    symbolic x coords={399114.52,45691.64,19753.31,14545.98,4096,3492.41,604.71},
    ymax=200,
    ylabel={\parbox{2cm}{\centering \textcolor{red}{Leaked bits/h [\_\_]} }},
    ymajorgrids=true,
    grid style=dashed,
    scaled x ticks = false,
    x tick label style={font=\normalsize, rotate=30, anchor=east},
    legend style={at={(0.68,1.5)}, anchor=north east, legend columns=2, font=\tiny}
]
\addplot[red] coordinates {(399114.52,120) (45691.64,79) (19753.31,62) (14545.98,10) (4096,1) (3492.41,1) (604.71,0)};
\addlegendentry{Leakage}

\addplot[green,dashed,pattern=north west lines, pattern color=green] coordinates {(4096,0) (4096,180) (14545.98,180) (14545.98,0)};
\addlegendentry{No observable leakage but still detected}
\end{axis}

\begin{axis}[
  width  = {\hsize},
  height = 3cm,
  axis y line*=right,
  axis x line=none,
  symbolic x coords={399114.52,45691.64,19753.31,14545.98,4096,3492.41,604.71},
  ymin=0, ymax=110,
  ylabel={\parbox{2cm}{\centering \textcolor{blue}{Probability of detection}}},
  legend style={at={(1.01,1.5)}, anchor=north east, legend columns=1, font=\tiny},
]
\addplot[blue,dotted,thick]
  coordinates{
    (399114.52,100) (45691.64,100) (19753.31,100) (14545.98,100) (4096,100) (3492.41,0) (604.71,0)
};

\addlegendentry{Probability of detection}

\end{axis}

\end{tikzpicture}
}
    }
    \caption{Reducing the branch accesses / iTLB accesses and the corresponding leakage rat.}
    \label{fig:watering_down}
\end{figure}

Next, we look at the performance overhead of our attack in the case where the attacker tries to get below the detection thresholds.
Getting below this threshold requires the attacker to significantly slow down the amplified Spectre attack.
Since the attacker cannot get rid of the cache eviction, the number of amplification iterations has to be reduced.
Consequently, if the number of amplification iterations is reduced, the more requests, \ie samples, are required to clearly distinguish between a cache hit and miss (\cf \cref{fig:success_over_ampl,fig:success_over_requests}).
We evaluate the best possible attacker in native code who only mistrains one branch.
By omitting amplification or by applying a small factor of $10$, we can reduce the number of retired branch instructions / iTLB accesses on our test devices to \SIx{604.71} and \SIx{3492.41}, respectively, which is in the ranges of an average script.
However, with the latter, we observe a leakage rate of \SI{1}{\bit/\hour}.
Thus, we set the threshold to \SIx{4096} and receive an average false positive rate of \SI{0.61}{\percent} on our tested devices.
\cref{fig:watering_down} illustrates the decrease in leakage if the attack degrades from an amplified Spectre attack to a sequential attack.
Using a non-amplified approach, about \SIx{250000} requests are required (\cref{sec:attack_eval}). %
We achieve a leakage of \SI{1}{\bit/\hour} in a local-network scenario.
Hence, as an additional security margin, we limit the number of subsequent requests per \worker to \SIx{10000} on the same machine.
If more than \SIx{10000} requests are issued, we redirect the request to a different machine.
Thus, we can still detect a slowed-down attack using our threshold-based approach.
Additionally, if an attacker can get below the threshold, the number of requests on the same \worker is limited.

We assume that there are no attacks running on the production system, thus we cannot measure the number of false negatives.
Our own attack is detected by the threshold, as well as the \SIx{13} Spectre samples provided by Kocher~\cite{Kocher2018mitigations}.
In addition, we evaluated and analyzed the new and larger Spectre-PHT gadgets generated by FastSpec~\cite{Tol2020FastSpec}.
The gadgets are based on the the \SIx{15} variants, and we observe that the generated gadgets are quite similar.
We evaluated \SIx{100} random gadgets from FastSpec and did not observe any false negatives with our detection. 
This is due to the fact that the mistraining for those gadgets is similar, leading to a similar range of branch accesses per iTLB access.
We also evaluated the detection on the Spectre JavaScript PoC from Röttger and Janc~\cite{Tsuro2021SpectreJS}. 
Even with the low amplification factor of 4000 used in this PoC, we reliably detect the attack ($n=500$, $\mu=$\SI{19253.73}).

\textit{Spectre-BTB, Spectre-RSB and Spectre-STL.}
In addition to Spectre-PHT we also run our performance counter analysis on the other Spectre variants exploiting the branch-target buffer (BTB), return-stack buffer (RSB) and store-to-load (STL) forwarding.
We create native code proof-of-concepts for these variants which execute each gadget \SIx{10000} times on a Xeon Silver 4208.
We ran the PoCs 500 times and collected the number of branch accesses and iTLB accesses.
The numbers for Spectre-BTB and RSP are on order of magnitued lower compared to Spectre-PHT ($\mu_{\bar{btb}}=423171.54$). 
However, they are still detected with the same metric ($n=500$): %
Spectre-BTB ($\mu_{\bar{btb}}=23401.20$), Spectre-RSB ($\mu_{\bar{rsb}}=38369.17$), Spectre-STL ($\mu_{\bar{stl}}=982.20$)).
The metric for Spectre-STL is far below the threshold of \SIx{4096}.
However, the performance counter values for \texttt{memory\_disambiguation.history\_reset} are significantly higher on average if the store-to-load logic is exploited in Spectre-STL ($n=500$, $\mu_{\bar{stl}}=8993.98$, $\mu_{\bar{nostl}}=2644.73$).
Thus, we additionally use this counter to detect potential Spectre-STL attacks. 

\subsection{\DefenseName}\label{sec:dynamic_isolation}

We integrate \DefenseName in \CFWorkers, which requires modifications of 6459 lines of code, not including the Spectre detection mechanism.
As with any isolation technology, the performance overhead will vary depending on the workload~\cite{Reis2018siteisolation}.
\CFWorkers is an environment where typical guest workloads use very little memory and spend very little CPU time responding to any particular event.
As a result, in this environment, dynamic process isolation overhead is expected to be large compared to the underlying workload.

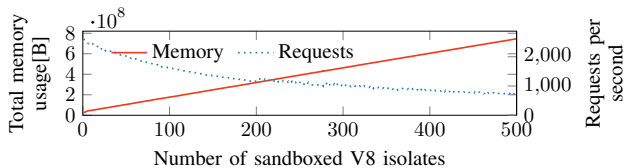
\begin{figure}[t]
    \centering
    \resizebox{\hsize}{!}{
            \resizebox{\hsize}{!}{
\begin{tikzpicture}
            \begin{axis}[
            width={\hsize},
            height=3cm,
            xlabel={Number of sandboxed V8 isolates},
            ylabel={\parbox{2.5cm}{\centering Total memory\\usage[B]}},
            xmin=0,
            ymin=0,
            xmax=500,
            legend style={at={(0.35,0.95)}},
            legend style={draw=none,fill=none}
            ]
            \addplot[color=red,thick]  table[x index = {0}, y index = {3}, col sep=comma]{data/result_isolation.csv}; \addlegendentry{Memory}
            \end{axis}
            
             \begin{axis}[
            width={\hsize},
            height=3cm,
            xlabel={Number of sandboxed V8 isolates},
            ylabel={\parbox{2cm}{\centering Requests per second}},
            xmin=0,
            xmax=500,
            ymin=0,
            axis y line*=right,
            axis x line=none,
            legend style={at={(0.65,0.95)}},
            legend style={draw=none,fill=none}
            ]
            \addplot[color=blue,dotted,thick]  table[x index = {0}, y index = {1}, col sep=comma]{data/result_isolation.csv}; \addlegendentry{Requests}
            \end{axis}

\end{tikzpicture}
}
    }
\caption{Performance overhead in terms of requests per second and total memory consumption of process isolation}
\label{fig:dynamic_isolation_overhead}
\end{figure}

In a first test, we evaluate the overhead for a test script by increasing the number of isolated processes, \ie the number of sandboxed V8 isolates, up to \SIx{500}.
We measure the overhead in terms of executed scripts per second, \ie the requests executed per second from the localhost and the total amount of consumed main memory.
The execution is repeated \SIx{10} times per isolation level with \SIx{2000} requests ($n=20000$, $\sigma_{\bar{rps}}=3.87\%$, $\sigma_{\bar{mem}}=0.23\%$).
\cref{fig:dynamic_isolation_overhead} shows the requests per second and the total memory consumption based on the number of isolated V8 processes.
As expected, we observe a linear decrease in the possible number of requests per second and a linear increase in the memory consumption.

Further, we performed a load test of \CFWorkers runtime using a selection of sample guest \workers simulating a heavy-load machine.
They mostly respond to I/O in under a millisecond and allocate little memory.
By forcing process isolation on the \workers, the memory overhead of each guest was 2x-5x higher, and CPU time was 8x higher, compared to a \worker using a single process.
We performed a second test using a real-world \worker known to be unusually resource hungry in both CPU and memory usage.
In this case, we found memory overhead to be \SI{20}{\percent}-\SI{70}{\percent} worse with dynamic process isolation, and CPU time to be about \SI{60}{\percent} worse.

These numbers appear to be high, but when only \SI{0.61}{\percent} of workers are isolated, the overhead is negligible.
As our proof of concept was not optimized, it still has big potential for optimizations.
For example, it currently uses an RPC protocol~\cite{Varda2019CapnProtoAnon} to communicate between processes, but does so over a Unix domain socket.
This protocol is designed in such a way that it could be communicated in shared memory, reducing communication overhead.
The implementation could also use OS primitives for faster context switching, such as the FUTEX\_SWAP feature proposed by Google~\cite{LKMLFutexSwap}.
However, while especially the CPU overhead could be reduced, there will always be significant cost incurred by context switching and marshalling needed to communicate between processes.
The total overhead on all machines can only be estimated as it depends on the workload.
The detection overhead is \SI{2}{\percent}.
In the worst case, we are slightly worse than full process isolation due to the additional \SI{2}{\percent} overhead for the detection.
However, due to the low false-positive rate, this is extremely unlikely.

\section{Discussion}\label{sec:discussion}

\paragrabf{Mitigation versus Detection.}
The current mitigations for Spectre are costly~\cite{Canella2019A,Kocher2019,IntelMitigations}.
Especially in high-performance scenarios, such as cloud systems, the mitigations result in high power consumption.
Hence, instead of paying the constant costs of mitigations, it can be desirable to reduce the costs by relying on the detection of attacks.
However, the problem of detecting side-channel and transient-execution attacks is still an open research problem.
There is no universal solution that covers all different types of attacks.

\paragrabf{False Positives.}
Our detection relies on performance counters, and, thus, false positives will occur~\cite{Das2019Sok,Zhou2018Hardware}.
The advantage of \DefenseName is that the detection mechanism can tolerate false positives.
In comparison to machine learning approaches that try to reduce the false-positive rate~\cite{Li2018online,Gulmezoglu2019}, we propose a simple threshold to detect potential Spectre attacks with a false-positive rate of \SI{0.61}{\percent}.
This approach is lightweight, easy to integrate into production software, and has a small performance overhead.
Our approach can be extended to more sophisticated machine learning approaches as proposed in related work~\cite{Li2018online,Gulmezoglu2019} to reduce the false-positive rate, as long as the false-negative rate does not increase.

\paragrabf{False Negatives.}
Our detection relies on performance counters, and, thus, false negatives will occur~\cite{Das2019Sok,Zhou2018Hardware}.
This is similar to other detection and mitigation techniques, such as oo7~\cite{Wang2019oo7} or Spectector~\cite{Guarnieri2020spectector}, which suffer from false negatives for novel forms of Spectre gadgets~\cite{Tol2020FastSpec}.
Nevertheless, with the current limits given in~\cref{tab:current_limits} (\Cref{sec:appendix:limits}), the attack would always exceed the CPU runtime constraint. 
As \CFWorkers considers increasing the limits in terms of execution time, an attack could be feasible.
Therefore we evaluated the attack in terms of the best possible attack we found.
The results of~\cref{tab:js_attack} (\Cref{sec:appendix:limits}) show that it is possible to get below the threshold, leading to false negatives.
But with an amplification factor of $10$, the leakage rate is only about \SI{1}{\bit/\hour} and requires about \SIx{25000} requests.
By limiting the number of subrequests to \SIx{10000}, we mitigate this special case where the attacker slows down the attack.

\begin{figure}[t]
  \centering
  \resizebox{\hsize}{!}{
  \resizebox{\hsize}{!}{
\begin{tikzpicture}[transform shape]
\begin{axis}[
    ylabel={normalized factor},
    xlabel={additional code pages},
    width  = {\hsize},
    height = 3cm,
    ytick={1,1e-2,1e-4},
    xtick=data,
    ymode=log,
    legend pos=north east,
    ymajorgrids=true,
    grid style=dashed,
    scaled x ticks = false,
    legend style={anchor=west}
]

\addplot+ [] table [col sep=comma,x=pages,y=factorn] {data/factor_over_pages.csv};
\end{axis}

\end{tikzpicture}
}
  }
  \caption{Influence of the number of additional code pages on the branch accesses / iTLB accesses.}
  \label{fig:factor_over_pages}
\end{figure}
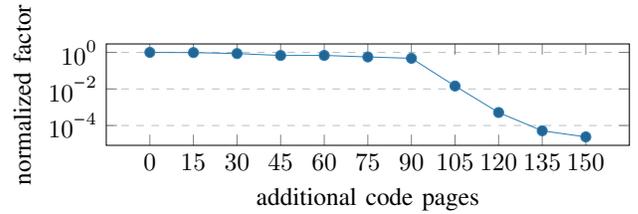

Adding additional code pages might allow an attacker to get below the proposed thresholds. 
We evaluated the influence of the number of additional code pages on the detection factor in \Cref{fig:factor_over_pages} (\Cref{sec:appendix:limits}).
To ``hide'' the native attack, we access 125 additional code pages per bit to get the branch accesses / iTLB accesses below the threshold. 
In total, this adds \SI{500}{\kilo\byte} code.
While feasible in native code, the JavaScript environment is more constrained. 
We tried different methods to inflate the binary size. 
In all cases, the resulting code size causes V8 to abort the optimization phase, stopping the attack.

\paragrabf{PEBS-based Detection.}
PEBS-based detection mechanisms seem to be an interesting research topic for future work, especially with the precise timestamps provided by PEBS.
However, while working with PEBS, we discovered a bug in the Linux perf implementation, where Linux accidentally overwrites the precise timestamp.
We reported this bug and submitted a patch to LKML~\cite{Cloudflare2019Anon}.

\paragrabf{Reliability of HPCs In \DefenseName}
As Zhou~\etal\cite{Zhou2018Hardware} and Das~\etal\cite{Das2019Sok} discuss, using HPCs for detection of microarchitectural attacks can lead to flaws caused by non-determinism and overcounting.
Our approach uses the \texttt{perf} interface to setup the HPCs.
However, by leveraging the \texttt{rdpmc} instruction, the counter values are read out directly from the model-specific registers.
We consider non-determinism and overcounting effects by averaging multiple times over the performance counter results for each script.

\paragrabf{Alternative Spectre JS attacks}
Concurrent work~\cite{Tsuro2021SpectreJS} has demonstrated an amplifiable Spectre exploit on V8, able to leak up to \SI{60}{B/s} using timers with a precision of \SI{1}{\milli\second} or worse through a side channel on the L1 cache. 
Similarly to our PoC, it uses a Spectre-PHT gadget to read out-of-bound from a JavaScript TypedArray, giving an attacker access to the entire address space.
The PoC uses small-sized TypedArrays for which the backing store is allocated in the isolate itself.
Thus, it leaks data inside the same isolate. 
During our evaluation, we extended the published PoC with two additional steps. 
We first leverage the leaks inside the isolate to disambiguate the layout of the neighboring objects and locating the address of a controlled array. 
Then we leverage a custom Spectre V1 gadget to perform a \emph{speculative type confusion} between two large objects that span multiple cache lines so that the type information of both objects are on different cache lines than the field target of the type confusion.
Speculating past the type check dereferences an attacker-provided integer. 
Thus, the Spectre gadget can load data from an arbitrary 64-bit address.
However, our enhanced PoC requires two memory dereferences during the speculation window.
We found they can be completed reliably only if the type information of the offending object is not cached while the integer containing the address is cached. %
Evicting only the type information from the cache requires an eviction set, and we failed to obtain that using only a remote timer.

\section{Conclusion}\label{sec:conclusion}

In this paper, we presented \DefenseName, a practical low-overhead solution to actively detect and mitigate Spectre attacks.
We first presented an amplified JavaScript remote attack on \CFWorkers, which leaks \SI{120}{\bit/\hour}.
We proposed a novel approach of actively detecting Spectre attacks by sampling mispredicted and retired branches, which achieves a false positive rate of \SI{2.83}{\percent}.
However, due to the large number of samples required, it leads to a total overhead of \SI{50}{\percent} on the production environment.
We show that it is still possible to efficiently detect Spectre attacks using performance counters with a false-positive rate of \SI{0.61}{\percent} at the cost of \SI{2}{\percent} overhead for the detection. %
We demonstrate that conditionally applying process isolation based on a detection mechanism has a better performance than full process isolation, while still providing the security guarantees of process isolation.

\bibliographystyle{plainurl}
\bibliography{main}

\FloatBarrier

\appendices

\section{Current limits of default \worker and JS attack numbers}
\label{sec:appendix:limits}
The state-of-the art limits of the \CFWorkers setup for a single instance is given in~\cref{tab:current_limits}.
We provide the leakage rate depending on the amplification factor in~\cref{tab:js_attack}.
\begin{table}[t]
  \caption{Current limits of \worker.}
\vspace{-0.1cm}
\setlength{\aboverulesep}{0pt}
\setlength{\belowrulesep}{0pt}
    \begin{center}
      \adjustbox{max width=\hsize}{
      \begin{tabular}{ll}
        \toprule
        \textbf{Configuration}                    &  Value \\
        \midrule
        Allowed Consumed Memory                   & \SI{64}{\mega\byte}  \\
        Compile Time                              & \SI{200}{\milli\second}  \\
        CPU runtime per request                   & \SI{50}{\milli\second}  \\
        \# of subrequests per request             & \SI{50}{requests} \\
        \bottomrule
      \end{tabular}
    }
  \end{center}
  \label{tab:current_limits}
\end{table}

\begin{table}[t]
  \caption{Attack results of our JS attack.}
\vspace{-0.1cm}
\setlength{\aboverulesep}{0pt}
\setlength{\belowrulesep}{0pt}
    \begin{center}
      \resizebox{\hsize}{!}{
      \begin{tabular}{rrrr}
        \toprule
        \textbf{Amplification}             &  Required requests & Script runtime & Leaked bits/hour \\
        \midrule
        \SIx{1}                           & \SIx{250000}     & \SI{118}{\milli\second}    & \SI{0}{\bit/\hour} \\
        \SIx{10}                          & \SIx{25000}      & \SI{123}{\milli\second}    & \SI{1}{\bit/\hour} \\
        \SIx{100}                         & \SIx{2500}       & \SI{137}{\milli\second}    & \SI{10}{\bit/\hour} \\
        \SIx{1000}                        & \SIx{250}        & \SI{231}{\milli\second}    & \SI{62}{\bit/\hour} \\
        \SIx{10000}                       & \SIx{25}         & \SI{1813}{\milli\second}   & \SI{79}{\bit/\hour} \\
        \SIx{250000}                      & \SIx{1}          & \SI{30000}{\milli\second}  & \SI{120}{\bit/\hour} \\
        \bottomrule
      \end{tabular}
    }
  \end{center}
  \label{tab:js_attack}
\end{table}

\section{Optimizations of Spectre attack}
\label{sec:appendix:optimization_prevention}

The function containing the Spectre gadget accesses the attacker's {\tt ArrayBuffer} through differently-sized {\tt TypedArray}s.
This prevents the JIT compiler from making assumptions on the memory accesses on the {\tt ArrayBuffer}.
Otherwise, the JIT compiler hard-codes the size for the bounds check, which significantly decreases the success probability for the Spectre attack.
As the garbage collector moves objects around, using multiple {\tt TypedArray}s increases the probability of having correctly aligned objects, such that the backing store pointer and the size of the {\tt ArrayBuffer} lie on different cache lines.

We specifically avoid triggering any de-optimization points in our generated code (\ie breaking assumptions of the JIT compiler), as that ruins the predictor's training.
Therefore, we place the out-of-bound access of the {\tt TypedArray} behind a mispredicted guard branch that depends on a variable with a value strictly lower than the size.
This prevents the JIT compiler from de-optimizing the code when detecting out-of-bound accesses.
We additionally increase the size of the attacker function to avoid it being inlined, which would cause de-optimization if the calling function is de-optimized, and would prevent training the predictors if called from different call sites.
The function takes the offset to access as a parameter and is called in a loop with a branch-less code sequence that feeds it with four in-bound offsets and an out-of-bound offset to access.

We warm-up the JIT compilation by repeatedly calling our function using an out-of-bound index that is higher than the one in the guard branch, but in-bound with respect to the {\tt TypedArray}, preventing the JIT compiler from making assumptions on the provided value.
We additionally found that executing a different number of taken conditional branches before executing the target function affected the leakage rate considerably.
The taken branches seem to affect code alignment and predictor states.
By automatically tuning the number of such branches, we empirically verified that \SI{70} is the optimal number for our PoC.

\subsection{Other Detection Methodologies}\label{sec:other_methodologies}
On Skylake and newer generations, PEBS provides a precise timestamp of the sampled events~\cite{Intel2017Performance}.
As the \texttt{BR\_INST\_RETIRED.NOT\_TAKEN} event does not support PEBS, we use the conditional mispredicted branches (\texttt{BR\_MISP\_RETIRED.CONDITIONAL}) and retired near taken call (\texttt{BR\_INST\_RETIRED.NEAR\_TAKEN}) events.
By sampling the timestamps, it is possible to timely interleave the two events.
As a result, it is possible to record a Spectre-PHT in-place exploitation pattern, \ie multiple correct predictions and then one misprediction.
Using this setup, we successfully detect a Spectre-PHT in-place exploitation pattern.
While this approach detects a Spectre-PHT attack, it has a high false-positive rate, as other structures result in a similar pattern, such as, \eg loops.
Moreover, PEBS with precise timestamps is only available since Skylake, which limits its applicability in real-world scenarios.

Another approach proposed to detect Spectre-attack patterns is to use the Last Branch Record (LBR)~\cite{ElasticBlog2018}.
While this approach counts how often a basic block is executed, it does not provide information on whether a branch to a basic block was correctly predicted.
Furthermore, we found that the resolution of the LBR (32 entries on Skylake) is too small to reliably detect attacks.
Alternatively, the Branch Trace Store (BTS) collects exact data on branch mispredictions~\cite{Intel2017Performance}, but introduces large performance overheads of up to \SI{40}{\percent}~\cite{Soffa2011Exploiting}.

\section{Time Shifting in V8}\label{sec:appendix:timeshift}
We observe a time shifting of the time stamp counter when leaking several bits after each other.
After a random amount of time, the execution time of the function shifts to another range.
We assume that some parts of the V8 engine might cause this effect, which we leave for future work to analyze.

\section{Phoronix Benchmarks}\label{sec:appendix:phoronix}
\cref{tab:evaluated_programs} lists all evaluated programs from the Phoronix benchmark suite for the PEBS-based approach (\cf \Cref{sec:eval_histogram}) as well as the \SIx{13} Spectre gadgets from Kocher~\cite{Kocher2018mitigations}.

\onecolumn

\begin{table}[H]
  \caption{Evaluated programs of the Phoronix benchmark suite~\cite{Phoronix2020Benchmark}.}
\vspace{-0.1cm}
\setlength{\aboverulesep}{0pt}
\setlength{\belowrulesep}{0pt}
\begin{center}
\resizebox{\hsize}{!}{
    \begin{tabular}{|c|c|c|c|c|c|}
        \hline
        System Tests        & Disk Tests & \multicolumn{3}{c}{Processor Tests} & Kocher Tests                                                  \\
        \hline
        apache              & aio-stress & aircrack-ng                         & gmpbench              & noise-level             & Spectre V01 \\
        caffe               & sqlite     & amg                                 & gnupg                 & npb                     & Spectre V02 \\
        compress-lzma       &            & aobench                             & go-benchmark          & n-queens                & Spectre V03 \\
        compress-pbzip2     &            & aom-av1                             & gpu-residency         & oidn                    & Spectre V04 \\
        compress-zstd       &            & blosc                               & graphics-magick       & openssl                 & Spectre V07 \\
        gnupg               &            & bork                                & hackbench             & openvkl                 & Spectre V08 \\
        openssl             &            & botan                               & himeno                & ospray                  & Spectre V09 \\
        redis               &            & bullet                              & hmmer                 & parboil                 & Spectre V10 \\
        battery-power-usage &            & byte                                & hpcg                  & perl-benchmark          & Spectre V11 \\
                            &            & cachebench                          & ipc-benchmark         & polybench-c             & Spectre V12 \\
                            &            & clomp                               & java-gradle-perf      & polyhedron              & Spectre V13 \\
                            &            & cloverleaf                          & java-scimark2         & povray                  & Spectre V14 \\
                            &            & compress-7zip                       & john-the-ripper       & primesieve              & Spectre V15 \\
                            &            & compress-gzip                       & lammps                & psstop                  & FastSpec gadgets \\
                            &            & compress-xz                         & lczero                & qmcpack                 &             \\
                            &            & core-latency                        & libgav1               & radiance                &             \\
                            &            & coremark                            & llvm-test-suite       & rays1bench              &             \\
                            &            & cp2k                                & luajit                & redis                   &             \\
                            &            & cpuminer-opt                        & lulesh                & renaissance             &             \\
                            &            & crafty                              & luxcorerender         & rodinia                 &             \\
                            &            & c-ray                               & mafft                 & scimark2                &             \\
                            &            & cython-bench                        & mencoder              & smallpt                 &             \\
                            &            & dacapobench                         & minife                & sqlite                  &             \\
                            &            & dav1d                               & minion                & stockfish               &             \\
                            &            & dcraw                               & mkl-dnn               & sudoku                  &             \\
                            &            & deepspeech                          & mpcbench              & svt-av1                 &             \\
                            &            & dolfyn                              & m-queens              & svt-hevc                &             \\
                            &            & ebizzy                              & mrbayes               & svt-vp9                 &             \\
                            &            & embree                              & java-gradle-perf      & swet                    &             \\
                            &            & encode-flac                         & java-scimark2         & system-decompress-bzip2 &             \\
                            &            & encode-mp3                          & john-the-ripper       & system-decompress-tiff  &             \\
                            &            & encode-wavpack                      & mt-dgemm              & system-decompress-xz    &             \\
                            &            & espeak                              & multichase            & system-libjpeg          &             \\
                            &            & ffmpeg                              & namd                  & system-libxml2          &             \\
                            &            & ffte                                & neat                  & tachyon                 &             \\
                            &            & fftw                                & nero2d                & toybrot                 &             \\
                            &            & fhourstones                         & nettle                & tscp                    &             \\
                            &            & glibc-bench                         & node-express-loadtest & ttsiod-renderer         &             \\
                            &            & tachyon                             & node-octane           & tungsten                &             \\
                            &            & toybrot                             & ttsiod-renderer       & vpxenc                  &             \\
                            &            & tscp                                & x264                  & x265                    &             \\
                            &            & yafaray                             & y-cruncher            &                         &             \\
        \hline
    \end{tabular}
}
    \end{center}
  \label{tab:evaluated_programs}
\end{table}

\twocolumn

\end{document}